\documentclass[12pt,a4paper]{article}%
\pdfoutput=1

\usepackage[T1]{fontenc}
\usepackage[utf8]{inputenc}
\usepackage{amsmath}
\usepackage{amssymb}
\usepackage{amsthm}
\usepackage[ 
    ]{hyperref}
\usepackage{mathtools}
\usepackage{graphicx}
\usepackage{booktabs}
\usepackage{caption}
\usepackage{slashed}
\usepackage{listings}
\usepackage{slashed}
\usepackage{rotating}
\usepackage{subfig}
\usepackage{geometry}
\usepackage{multicol}
\usepackage[nottoc]{tocbibind}
\usepackage{multicol}
\usepackage{authblk}
\usepackage{cite}
\usepackage{appendix}
\usepackage{color}
\usepackage{epsfig}
\usepackage{hyperref}
\usepackage{enumerate}
\usepackage{amsfonts}
\usepackage{epstopdf}
\usepackage{latexsym}
\usepackage{arydshln}

\definecolor{myred}{rgb}{0.7, 0, 0}
\definecolor{myblue}{rgb}{0, 0, 0.7}
\definecolor{mygreen}{rgb}{0.04, 0.7, 0.5}

\hypersetup{colorlinks,citecolor=myred,linkcolor=myblue,urlcolor=myblue,linktocpage=true}

\newcommand{\be}{\begin{equation}}
\newcommand{\ee}{\end{equation}}
\newcommand{\bea}{\begin{eqnarray}}
\newcommand{\eea}{\end{eqnarray}}
\newcommand{\eq}[1]{Eq.~(\ref{#1})}

\newcommand{\lc}{\Lambda_{ c}}
\newcommand{\luv}{\Lambda_{\rm UV}}
\newcommand{\lssb}{\Lambda_{\rm SSB}}
\newcommand{\fssb}{f_{\rm SSB}}

\newcommand{\zir}{z_{\rm IR}}
\newcommand{\lqcd}{\Lambda_{\rm QCD}}
\newcommand{\lqcddec}{\Lambda_{\rm QCD}^{\rm dec}}
\newcommand{\I}{\ref{typeI} }
\newcommand{\II}{\ref{typeII} }
\newcommand{\Ico}{\ref{typeI}, }
\newcommand{\IIco}{\ref{typeII}, }

\numberwithin{equation}{section}

\title{\Large \textbf{
The  Supercooled   Universe}}

\author{
{\large Pietro Baratella$^{a}$, Alex Pomarol$^{a,b}$ and Fabrizio Rompineve$^{a}$}\\
\normalsize\itshape $^a$ IFAE and BIST, Universitat Aut\`onoma de Barcelona, 08193~Bellaterra,~Barcelona\\
\normalsize\itshape $^b$ Dept.~de~F\'isica, Universitat Aut{\`o}noma de Barcelona, 08193~Bellaterra,~Barcelona\\
}

\begin{document}

\maketitle

\begin{abstract}

\noindent   Strongly-coupled theories at the TeV can naturally  drive   a long period of  supercooling
in the early universe.  Trapped into  the deconfined phase,  the universe   could inflate and cool down 
till the temperature reaches  the  QCD strong scale. We show how at these
 low temperatures  QCD effects are important and could  trigger the  exit from the long supercooling era. 
We also study  the implications   on  relic abundances.
 In particular,  the latent heat released at the end of supercooling could be the reason
 for the similarities between dark matter and baryon energy densities.
The axion abundance could also be significantly  affected,
 allowing for  larger values of the axion decay constant.
Finally, we  discuss how  a long supercooling epoch 
could lead to an enhanced  gravitational wave signal.

\end{abstract}

\newpage

\section{Introduction}

As  the LHC continues exploring the TeV-scale  territory,  searching for  hints on the origin of the electroweak (EW) scale, 
it is also quite  necessary to look for probes of new  physics  in other  environments, as for example  in the early universe.
This is of   vital importance  due  specially to the  LHC energy limitation,  
as  these probes could indirectly provide evidences for new physics just beyond the LHC reach.

New strongly-interacting sectors at the TeV scale are  particularly  well-motivated examples 
 for explaining the origin of the EW scale,
and are at present  exhaustively searched at the LHC.
These models also present dark matter (DM) candidates that are   searched  for at present experiments.
In spite of this, there is  a generic prediction  for  these  TeV strong sectors 
that  has not been fully explored  and
can have a strong  impact  in the physics of the early universe:
 a  long epoch of supercooling   arising  as a result of the  transition from the deconfined to the confined phase.
The main purpose of this article is to  address   under which conditions this long epoch of supercooling can happen,
and  explore the  phenomenological  consequences  in the early universe.

The basic dynamics which gives birth to such a supercooled universe is  the following.
 Strongly-interacting theories  are expected to be 
 in a deconfined phase  at temperatures above the TeV, in which its constituents behave as a strongly-interacting hot gas. 
When the temperature goes down below the TeV,   the  confined phase becomes  energetically favorable.
Nevertheless, when the  rank of the strong gauge group $N$  is large,
 the phase transition is of first-order and can only be completed through tunneling. 
Since the tunneling rate is quite suppressed $\Gamma\propto e^{-S_B}\sim e^{-N^2}$,  the transition cannot occur,
and the universe enters a potentially eternal epoch of inflation trapped in the deconfined phase.
As the temperature decreases, however,  the QCD critical temperature is attained. 
Since  these TeV strong sectors must be coupled to the  SM quarks (to give them masses after the EW breaking),
strong QCD effects become important at low temperatures and can be responsible for 
triggering the phase transition and  ending the supercooling era.
Therefore,  a natural consequence of TeV-scale strong sectors is a  long epoch of  supercooling,  which can
last for several efolds.

The idea of a supercooled universe in which  the phase transition is driven by  QCD condensation
 was first considered in~\cite{Witten:1980ez}, in the context of the 
Coleman-Weinberg model, and  has been also recently discussed in~\cite{vonHarling:2017yew}
in strongly-coupled models similar to the ones discussed here.\footnote{Supercooling  with holographic duals was  first  considered in~\cite{Creminelli:2001th}.
  Since then, several studies of the confinement phase transition have been performed with short periods of supercooling~\cite{Randall:2006py, Kaplan:2006yi, Nardini:2007me, Hassanain:2007js, Konstandin:2010cd, Konstandin:2011ds, Konstandin:2011dr, Bunk:2017fic, Dillon:2017ctw, Megias:2018sxv}.}
Our first  main goal in this article is to  address  the general  conditions under which this long epoch occurs.
We will define the essential ingredients  of the strongly-coupled models needed for supercooling
and exiting at temperatures around or below the QCD strong scale.
As we will see, it  will be crucial to distinguish between models in which 
the EW symmetry is restored  in the  deconfined phase from those in which it is broken.
This latter possibility has not been considered  before, and, as we will see, can lead to important implications for axion physics.
We will also present holographic examples of these type of models that  allow to calculate tunneling rates
and determine the exit from supercooling.

Next, we will explore the consequences for cosmology.
We will see that  the  release of entropy at the end of the phase transition leads to  
a strong dilution  of   DM relics.
This implies a different range of couplings and masses for WIMP-like DM candidates and wimpzillas.
Also for DM candidates that  get their masses after the phase transition (e.g., resonances of the strong sector), we will see that their relic abundances  are mainly determined by their initial thermal abundance and the dilution from supercooling.
Interestingly, we will show that this can also be the case for the baryon asymmetry, and therefore the supercooled universe could give an explanation for the observed similarity  between 
 the baryon  and DM energy density.

We will also study the implications for the QCD axion.
We will see that during the supercooling epoch the axion could oscillate  and quickly relax to the minimum of its potential. Therefore, its relic abundance  could be reduced, meaning that larger values of the axion decay constant $F_a$ will be 
favored for  axion DM.
Whenever  axionic topological defects are relevant, their contribution to the axion relic abundance can also be enhanced in the supercooled universe. This is essentially due to the fact that the inflationary expansion delays the collapse of the axionic string-wall network compared to the standard case.

Perhaps the most model-independent implication of a confinement phase transition is the  gravitational wave (GW) signal. 
This has received a lot of attention in recent literature, but not for the particular   case of long supercooling  in which the thermal plasma which surrounds the colliding bubbles of the confined phase is very diluted. 
We will see  that the GW signal after a long supercooling epoch can be  sourced mostly by the collisions of bubbles of the confined phase, which occurs on a long time scale and thus maximizes the amplitude of the signal.

The organization of the paper is as follows.
In Section~\ref{SIS} we introduce the properties of the strongly-coupled sectors 
 in the confined phase and deconfined phase (Sec.~\ref{sec:dec}).
In Section~\ref{sec:rate} we analyze the deconfinement-confinement phase transition, calculating the tunneling rates
and the temperature for exiting.
In Section~\ref{sec:cosmology} we study the cosmological consequences of a long period of supercooling
for dark matter (Sec.~\ref{sub:dm}) and for baryons (Sec.~\ref{sub:baryons}).
Axion dark matter from misalignment is discussed  in Sec.~\ref{sub:misalignment},  while 
the impact on  axionic topological defects is given in Sec.~\ref{sub:topological}. 
Gravitational wave signals are discussed  in  Sec.~\ref{sec:gws}.
Section~\ref{conclusions} is devoted to conclusions.
We add two appendices. In Appendix~\ref{appendixA}
we present holographic models with the same properties of the strongly-coupled sectors that lead to supercooling.
In Appendix~\ref{app:topological}  we review  some basics of axion DM from the string-wall network.

\section{A strongly-interacting new sector at the TeV}
\label{SIS}

We are  interested in studying the supercooled phase of a   strongly-interacting sector, 
or its  equivalent five-dimensional  version  based on the AdS/CFT correspondence.
We will  identify here  the generic  properties (at zero $T$) of this class of theories  relevant for our analysis:
\begin{itemize}
\item
 We  assume that   these are strongly-interacting  conformal field theories (CFT) with a large central charge, or, equivalently, a large number of ``colors'' $N$.  This includes any gauge theory, e.g.  $SU(N)$,  in the  large $N$ limit.
 \item
These  CFTs  will contain a  small marginally relevant deformation that will be responsible for generating at the IR the confinement scale $\lc\sim$ TeV, much smaller than any UV-scale, $\luv$, as for example,  the Planck-scale. 
This deformation can arise from  a marginal operator added to the CFT, e.g.,
\be
\Delta {\cal L}=g\,  {\cal O}_g\,,
\label{deformation}
\ee
 where  Dim$[{\cal O}_g]=4-\epsilon$ with $\epsilon\ll  1$. The marginal coupling $g$
RG-evolves towards the IR  until a critical value $g_c$ is reached  at which  conformal invariance is  lost and 
a   mass-gap $\lc$ is generated. 
In this class of theories, like in QCD,  we expect a  tower of resonances of different spin, weakly-coupled in the large $N$ limit,
and with masses of order $\lc$. This   corresponds to the confined phase.
\item
As these models have usually   accidental global symmetries, $\cal G$,
an important parameter   is the scale of spontaneous symmetry breaking (SSB) $\cal G\to H$.
We will refer to this scale as $\lssb$.
This is associated to the nonzero condensate of an operator of the model ${\cal O}_H$ transforming under $\cal G$. For example, in QCD this is supposed to be  the  quark condensate $\langle \bar qq\rangle$. 
We will assume, like in QCD or other gauge theories studied in the lattice, that   $\lssb\sim \lc$ (all condensates are of the same order).
This is needed for phenomenological reasons, as  the scale
 $\lssb$ is associated with the  electroweak  scale that cannot be  larger than the masses of the new resonances,  bounded by the LHC to be above the TeV.
 For example, in  composite Higgs models \cite{Agashe:2004rs} where 
 the Higgs is a PGB arising from the breaking $\cal G\to H$,
 the Higgs VEV  is given by $v\simeq c\fssb$ where $c$ is an $O(1)$ coefficient and
 $  \fssb\simeq (\sqrt{N}/4\pi) \lssb$.
 In fact, in many models a small tuning is required ($c\lesssim 0.2$) 
 to fulfill  the experimental constraints \cite{Panico:2015jxa}. 
 \item
 In order to study the phase transitions of the above models, it is crucial to understand these theories off-shell.
This is a very difficult task, even for the holographic versions discussed in  Appendix~\ref{appendixA}.
Nevertheless,  this problem  simplifies substantially whenever the lightest state is the dilaton, $\mu(x)$, the 
Goldstone parametrizing the spontaneous breaking of the scale invariance.
In this case, we can study the effective potential of the dilaton after integrating out all other heavier states.
We can understand this potential based on symmetries.
Since under scale transformations, we have 
$x^\mu\to e^wx^\mu$ and $\mu\to e^{-w} \mu$,
the effective potential can be written as 
\be
V_{\rm eff}(\mu)=\frac{N^2}{16\pi^2} \lambda(\mu)\mu^4\,,
\label{potdilaton}
\ee
where the dependence of the quartic coupling $ \lambda(\mu)$  on $\mu$ is dictated by the explicit breaking of scale invariance.   We identify $\lc$ with the minimum of the dilaton potential \eq{potdilaton}, $\langle \mu\rangle=\lc$.
Notice that contrary to the Higgs $H$ and other fields, 
for the dilaton we will work in the basis in which its kinetic term is non-canonically normalized:
\be
\mathcal{L}_{\rm kin}(\mu)= c_1 \frac{N^2}{16\pi^2}(\partial_\mu\, \mu(x))^2\,,
\label{dilkin}
\ee
where $c_1$ is a constant of $O(1)$ and   we are assuming that the dilaton is a glueball.
This  normalization guarantees  $\langle \mu\rangle\sim \lc \sim O(N^0)$
and resonances get masses proportional to $\mu$ with no $N$ factors:
\be
m_i\simeq  r_i\langle\mu\rangle\,,
\label{massr}
\ee
where $r_i\sim O(1)$.
In holographic models  the dilaton is the radion and one obtains  $c_1=12$ and $r_i\sim \pi$  (see Appendix~\ref{appendixA}).
The mass of the dilaton is given by
\be
m_{\rm dil}^2=-\frac{2}{3}\left. \lambda(\langle \mu\rangle) \left(1+\frac{\beta_\lambda'}{4\beta_\lambda}\right)\right|_{\rm min}
\left(\frac{12}{c_1}\right)\langle\mu\rangle^2\,,
\label{mdil}
\ee
where $\beta_\lambda=d\lambda/d\ln\mu$  and $\beta_\lambda'=d\beta_\lambda/d\ln\mu$.
Notice that the dilaton effective potential makes sense for $m_{\rm dil}\lesssim m_i$; using \eq{massr}, this gives the rough bound  $|\lambda| \lesssim r^2_i\simeq  \pi^2$, where we have used holography in the last equality.\footnote{It is instructive to derive the value of $\lambda$ at the minimum
by equating \eq{potdilaton} to  the  QCD vacuum energy with $m_i \sim\pi\langle\mu\rangle\sim$ GeV and $N=3$. One obtains  
$\lambda_{\rm min}\simeq  0.9\, \pi^2$ as expected since QCD has no light dilaton.}
We will assume here that \eq{potdilaton} describes the  vacuum of the theory with the dilaton 
being the lightest state.
\item
An important property of this class of  models  that will be very relevant for the impact in cosmology
is  how the condensate ${\cal O}_H$, that breaks $\cal G\to H$, behaves
for  small  (off-shell) values of the dilaton $\mu$.
We will consider two possibilities:
\begin{enumerate}[ I.]
\item   \label{typeI}  All condensates are proportional to $\mu$. 
This implies in particular $\langle H\rangle\propto\mu$.
\item \label{typeII}  The condensate of ${\cal O}_H$ is determined to be  around $\lssb$,
independently  
of $\mu$.
We have then  $\langle H\rangle\propto \lssb$  even when $\mu\ll\lssb$.
\end{enumerate}
It is important to understand the behavior of $V_{\rm eff}(\mu)$  for the two types of models. 
In models of type \I  we expect   $\lambda$ to evolve logarithmically due to \eq{deformation}:
\be
 \lambda(\mu) \sim f(g(\mu)) \sim  \lambda_0+\beta_\lambda \ln\frac{\mu}{\luv}+\cdots\,,
\label{logdep}
\ee
where $\beta_\lambda\sim \epsilon$.
At around  $ \lambda(\mu)$ crossing zero,  $V_{\rm eff}(\mu)$ develops a minimum {\it \`a la} Coleman-Weinberg.
This is illustrated in  Fig.~\ref{lambdaplot}.
For constant $\epsilon$ we expect
\be
\langle \mu\rangle \sim e^{-1/\beta_\lambda}\luv\,,
\label{hierarchy}
\ee
which for $\luv\sim M_P$ requires $\beta_\lambda\sim 1/35$ in order to generate $\lc\sim$ TeV.
Holographic weakly-coupled models with this property
have been constructed, and are usually referred to as   Goldberger-Wise models \cite{Goldberger:1999uk}.
In these particular models $ f(g(\mu))$ can be computed and is a quadratic function of $(\mu/\luv)^{\epsilon}$ (see e.g.~\cite{Creminelli:2001th}).

On the other hand, models of type \II correspond, for example, to models at which
the dimension of the operator ${\cal O}_H$ is not constant but, due to \eq{deformation}, RG-evolves  
till it becomes imaginary at some scale, signaling the breaking of the conformal symmetry with 
$\langle {\cal O}_H\rangle\sim \lssb$. 
They are referred to as walking models (see for example \cite{Gorbenko:2018ncu})
and the holographic versions correspond to 5D models with an AdS tachyon as described in  Appendix~\ref{appendixA}.
For these type of models  one finds that for $\mu\sim \lssb$  the quartic goes as\cite{Pomarol:2019aae}
\be
 \lambda(\mu) \simeq  \lambda_0+\beta_\lambda \ln^2\frac{\mu}{\lssb}\,,
\label{logdep2}
\ee
with  $\beta_\lambda\sim$ 1, and   $V_{\rm eff}(\mu)$ can have a minimum  at some scale $\lc \sim \lssb$.
For $\mu<\lssb$ 
the quartic becomes almost constant (or at most evolving like \eq{logdep})  as
the theory   can  be described by  another CFT (since  part of the original CFT has become massive at $\lssb$).
This  is  shown in Fig.~\ref{lambdaplot}.

\begin{figure}[t]
\centering
\includegraphics[width=0.5\textwidth]{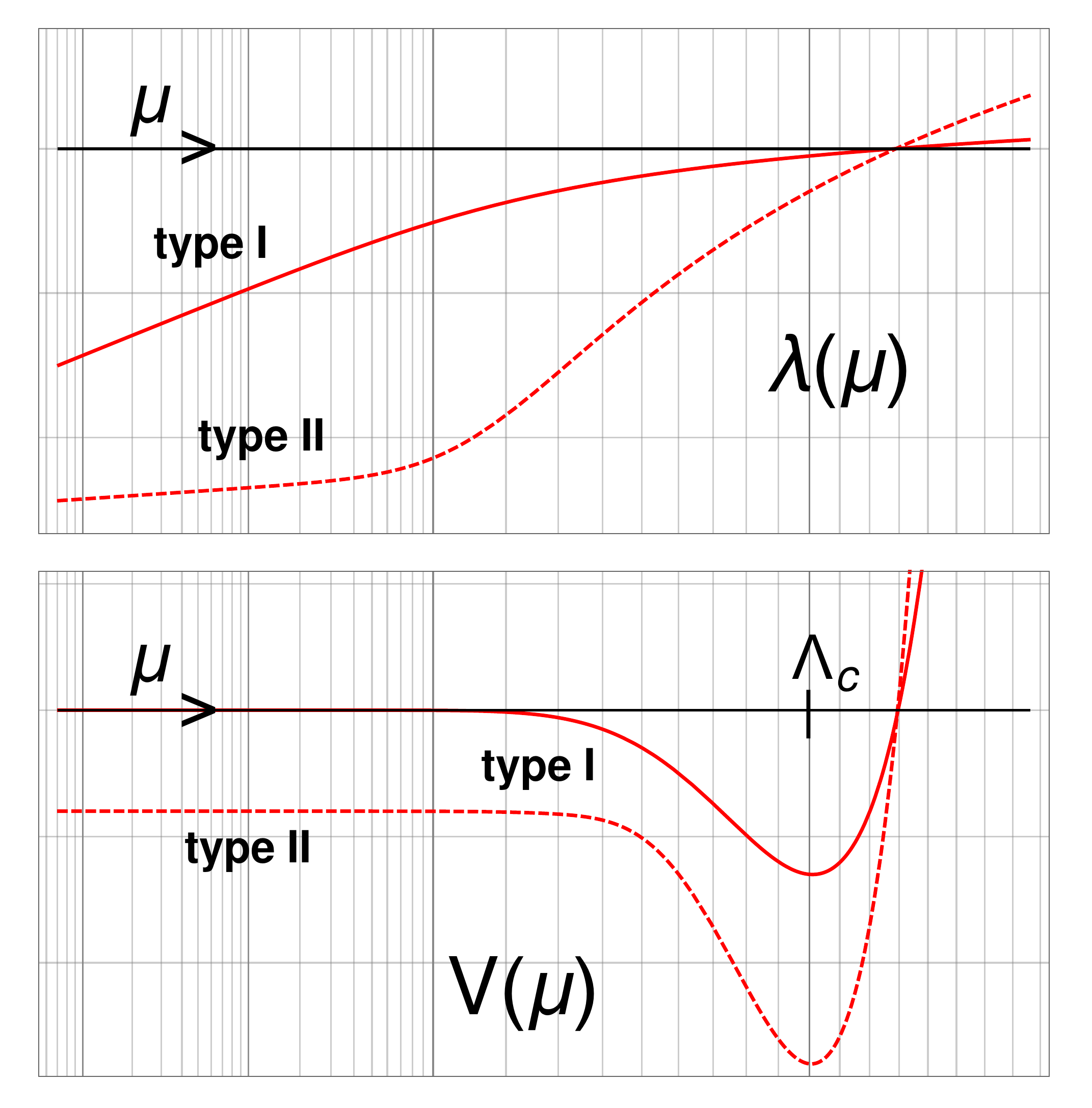}
\caption{\it  Behavior of the quartic coupling (upper plot) and effective potential for the dilaton (lower plot) as a function of $\mu$  for models of type \I  and type \II.}
\label{lambdaplot}
\end{figure}

\item
Another relevant ingredient of this class of models is how they couple to the SM quarks.
This coupling must be present,
since we need the Higgs (coming from the new strong sector) to give mass to the SM quarks.
The most realistic   possibility  is to  have a linear coupling of the SM quarks to  operators of the strong TeV sector:
\be
\Delta {\cal L}=\lambda_L\bar q_L\, {\cal O}_L+\lambda_L\bar q_R\, {\cal O}_R+h.c.\,,
\label{portal}
\ee
that  is the portal of the SM quarks to the  strong TeV sector, and therefore to the Higgs.
\eq{portal}  tells us that the composite operators ${\cal O}_{L,R}$ must transform as a fundamental representation under QCD,
that  implies that there must be  states charged under both the strong TeV sector and  QCD. 
This requirement will have important implications when, in the supercooling phase,  $T$  drops below
the QCD confinement scale $\lqcd$.
\end{itemize}
We can find weakly-coupled models with the properties described above by using the AdS/CFT correspondence 
(for models of type \Ico see \cite{Agashe:2004rs}, while for models of type \IIco see~\cite{Pomarol:2019aae}). 
The importance of these toy models is that they allow to perform calculations and  study the supercooling phenomena 
in a more quantitative way  (for previous work, see 
\cite{Creminelli:2001th,Randall:2006py, Kaplan:2006yi, Nardini:2007me, Hassanain:2007js, Konstandin:2010cd, Konstandin:2011ds, Konstandin:2011dr, Bunk:2017fic, Dillon:2017ctw, Megias:2018sxv}).

\subsection{Deconfined phase}\label{sec:dec}

At high-temperatures a strongly-interacting theory is expected to have a different more energetically favorable  phase
in which the constituents are not confined into hadrons. In QCD this is the  quark-gluon plasma phase.
For the models described above we will assume that indeed  this phase exists.

In this deconfined phase models of type \I and \II will have  different properties.
Models of type \I  are similar to QCD, in which  all   the condensates are zero and the theory has no mass gap  in the deconfined phase. In particular, we expect that the Higgs does not get a VEV in this phase.
On the other hand, in  models of type  \II we have that    
$\langle H\rangle\propto \Lambda_{\rm SSB}$ independently of the scale of condensation $\mu$, and therefore
the electroweak symmetry is  also broken in the deconfined phase, $\langle H\rangle_{\rm dec}\sim f_{\rm SSB}$, 
for $T< \Lambda_{\rm SSB}$.
We notice, as this will be important later when discussing axion physics, that $\langle H\rangle_{\rm dec}$ is expected to be slightly larger than the Higgs VEV in the confined phase, $\langle H\rangle\sim$ 246 GeV, since 
to  pass experimental constraints one  needs $f_{\rm SSB}\sim$  TeV \cite{Panico:2015jxa}. 

We will be interested later in studying the effects of  QCD condensation in the CFT deconfined phase.
 For this reason, it is important
to know  what is the  QCD scale in the deconfined phase $\lqcddec$. This scale  is generically different from its 
value in the SM vacuum $\lqcd\sim 300$ MeV as the CFT, which contains 
 charged states under QCD, can  contribute  to the running of $g_s$ below the TeV scale.
    The equation that governs the running of $g_s$ at the one-loop level    takes the well known form
\begin{equation}\label{gsrun}
\frac{d g_s}{d\ln Q}=\beta_{g_s}=\frac{g_s^3}{16\pi^2}\left(-\frac{11}{3}N_c+\frac{2}{3}N_f +\alpha N\right)\,,
\end{equation}
where $Q$ is the renormalization scale, while $N_f$ and $N_c$ are respectively the number of active flavors and colors. The contribution of the CFT is taken into account in the last term. Even though $\alpha$ is not computable, holography tells unambiguously that it is positive, since $\alpha N$ corresponds to  the inverse of the 5D gauge coupling squared in units of the AdS length \cite{Agashe:2004rs}.
By use of  \eq{gsrun} we find that
\be
\lqcddec = \lqcd \left( \frac{\lqcd}{\lc}\right)^{\frac{n}{1-n}},
\label{ldec}
\ee
where we normalized $\lqcddec$ to the SM value. The parameter $n$ is given by
\be
n=\frac{\beta_{g_s}(Q<\Lambda_c)-\beta_{g_s}(Q>\Lambda_c)}{\beta_{g_s}(Q<\Lambda_c)}\,.
\label{n}
\ee
\eq{ldec} is valid for $n<1$. For $n\geq 1$ the positive contributions to \eq{gsrun} win over the contributon of the gluons, and QCD is not confining, i.e., $\lqcddec=0$. 
Since the CFT always gives a positive contribution to the beta function, we have $\beta_{g_s}(Q>\Lambda_c)>\beta_{g_s}(Q<\Lambda_c)$,
\footnote{$\beta_{g_s}(Q<\Lambda_c)$ is negative because it is dominated by the contribution of the gluons.} and as a consequence $\lqcddec<\lqcd$. Nevertheless,  we can allow  for the possibility that the QCD $SU(3)_c$ group  arises as the low-energy remnant of a larger gauge  group ($SU(N_c)$ for simplicity), spontaneously broken by the strong dynamics of the CFT at the scale $\lc$. 
 In this case $n$ can be negative and then $\lqcddec>\lqcd$. 
 \footnote{The possibility to have a larger QCD scale in the early universe 
by giving  a VEV to a scalar  coupled to $G_{\mu\nu}^2$ has also been considered~\cite{Dvali:1995ce} (see also~\cite{Ipek:2018lhm}). This could also be possible here.}
 For example, for $\alpha N \sim 1$ and $N_c=5$, 
we get  $n=-0.5$,  giving  $\lqcddec \simeq 5 \,{\rm GeV} (\Lambda_c/{\rm TeV})^{1/2}$.

\section{Confined-deconfined phase transition and supercooling}\label{sec:rate}

If our universe was ever in the deconfined phase described in the previous section, it must have transitioned to the confined phase at some time during cosmic evolution. The temperature $T_c$
at which this becomes thermodynamically possible can be estimated as follows:  
in the unconfined phase the only relevant scale is temperature, and we then expect the free energy density to scale as in a hot free gas with $O(N^2)$ d.o.f.,  that is $\mathcal{F}_{\rm dec}\sim -N^2 T^4$. On the other hand,  in the confined phase the CFT acquires a mass gap, $\lc$, and at 
$T\ll \Lambda_{\rm c}$ the heavy resonances are not excited (due to Boltzmann suppression). Therefore the free energy is dominated by the temperature-independent vacuum energy, $\mathcal{F}_{\rm conf}\sim -N^2 \Lambda_{\rm c}^4$.
The phase transition can take place as soon as $T\lesssim T_c\sim \Lambda_{\rm c}$.

In large-$N$ gauge theories it is found that the confinement phase transition is of \textit{first order}; this is also seen in holographic models in which the two phases correspond to two different solutions of the same Lagrangian (see Appendix \ref{appendixA}), and therefore one can only go from one to the other by bubble nucleation. It is thus crucial to compute $\Gamma$, the rate at which bubbles are produced: in particular, the transition can be completed only when $\Gamma\gtrsim H^4$ (see e.g.~\cite{Gorbunov:2011zzc})
where $H$  is the Hubble constant that in the confined phase 
for $T<T_c$ is dominated by the vacuum energy:
\be
H\simeq \sqrt{\frac{|V_{\rm eff}(\langle\mu\rangle)|}{3M_P^2}}\simeq\frac{N}{4\sqrt{2}\pi}\frac{m_{\rm dil}\lc}{M_P}\,,
\label{hubble}
\ee
where we have used  \eq{mdil} with  $c_1=12$ and $\beta_\lambda'\lesssim \beta_\lambda$ as we will take from now on.
In a semi-classical approximation 
$\Gamma$ is given by
\begin{equation}
\Gamma= \mathcal{A}\,e^{-S_B}\,,
\end{equation}
where $S_B$ is the action of the critical bubble, or bounce, and $\mathcal{A}$ is typically of order $R^{-4}$, being $R$ the critical bubble radius. When the bounce is dominated by the dynamics of the large-$N$ sector,  we have 
$S_B\propto N^2$ and $\Gamma$ is exponentially suppressed. If the transition cannot take place the universe inflates in the confined phase and  exponentially cools down.

Before going into the details of the tunneling process, let us first compute more precisely the critical temperature $T_c$ below which confinement becomes favorable; for this we need to know the free energy in the two phases. As explained above, in the deconfined phase we can use conformal invariance and $N$-counting to fix $\mathcal{F}_{\rm dec}$ in the following way:
\be
\mathcal{F}_{\rm dec}= -c_2 N^2 T^4\,,
\label{fdec}
\ee
where $c_2$ is a constant of order 1; using holography we find $c_2=\pi^2/8$, an $O(10)$ larger than the free particle result (see Appendix~\ref{appendixA} for a derivation). In the confined phase, for temperatures below $\lc$, we have
\be
\mathcal{F}_{\rm con}=V_{\rm eff}(\langle\mu\rangle)+O(T^4)\,,
\label{fcon}
\ee
where thermal corrections mainly come from light degrees of freedom (always present in realistic models). The critical temperature is given by equating $\mathcal{F}$ in the two phases, which gives
\be
T_c\simeq 300 \,{\rm GeV}\left(\frac{\lc\, m_{\rm dil}}{\rm TeV^2}\right)^{1/2}\left(\frac{\pi^2 /8}{c_2}\right)^{1/4}\,,
\label{tcrit}
\ee
where we normalize $c_2$ with the value coming from holography.

\begin{figure}[t]
	\center
	\includegraphics[scale=.5]{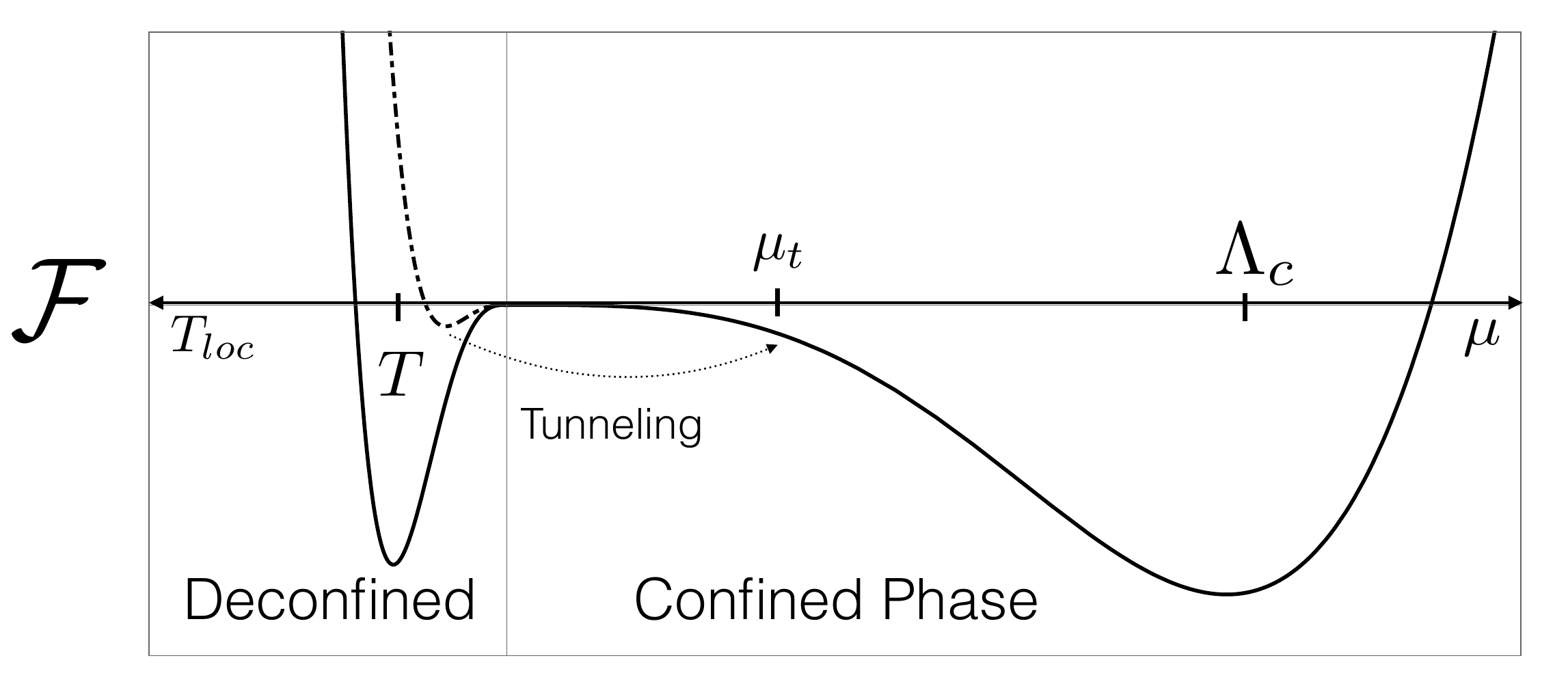}
	\caption{\it Free energy along the tunneling path described in the main text. In the deconfined phase (left side) it is a function of the local temperature $T_{loc}$, while in the confined phase it is a function of $\mu$. The two phases are identified at the point $T_{loc}=0=\mu$. On the left, we show the free energy profile for $T\lesssim T_c$ (solid line) and for $T\ll T_c$ (dot-dashed), to show in particular that a much larger region in $\mu$ space becomes energetically accessible as temperature decreases. On the right, we show the dilaton potential.
	}
\label{potential}
\end{figure}

When the temperature of the universe goes below $T_c$, bubbles of the confined phase begin to form.
It is in general not an easy task to compute the rate $\Gamma$ since  the bounce can depend on  many fields. To simplify the analysis, one usually looks for a preferred tunneling direction in field space, in order to  reduce it to a single-field problem. 
In the confined phase we already argued that the dilaton  is the only relevant field if we consider it lighter than the other resonances.
Nevertheless,   it  is not at all obvious what are the relevant field and potential in the deconfined phase, and how to properly connect the two phases. These points were first discussed by the authors of Ref.~\cite{Creminelli:2001th} in the context of the holographic 5D model. Here, we briefly rephrase their strategy in purely 4D terms. The key observation is that both phases 
 become indistinguishable in the limit $T=\mu=0$ at which  the conformal symmetry is recovered in both phases.
In Fig.~\ref{potential} we plot the potential along the path that connects in this way the deconfined phase (on the left side) to the confined phase (on the right side). 
In the confined phase the field variable is $\mu$, with potential $V_{\rm eff}$ given in \eq{potdilaton}, while in the deconfined phase the proper variable is the local temperature of the plasma, $T_{loc}$. 
The free energy density of the CFT plasma as a function of $T_{loc}$ (see left of Fig.~\ref{potential}) can be easily fixed, up to a constant factor, with elementary arguments from thermodynamics, and  has the following expression:
\begin{equation}\label{freeooe}
\mathcal{F}(T_{loc})=\rho(T_{loc}) -T\, s(T_{loc})=c_2 N^2\left(3 T_{loc}^4 - T\, 4T_{loc}^3 \right),
\end{equation}
where the constant is fixed in such a way that \eq{freeooe} matches with \eq{fdec} for $T_{loc}=T$. This result is also derived in Appendix \ref{appendixA} using  holography. To completely specify the system, we finally need to normalize the kinetic term for $T_{loc}(x)$. Even though it is not clear how to do it, not even from the holographic perspective, we take this to be $c_3 (N^2/16\pi^2)(\partial_\mu T_{loc})^2$. We will see however that  the parameter $c_3$ does not impact sensibly the  calculation of $\Gamma$.

At finite temperature bubble nucleation can be induced either thermally or quantum mechanically. Which process is more likely can be determined by comparing $T$ and $R^{-1}$: when $TR\gg 1$ (high temperature regime) thermal bubbles are preferred, while for $TR\ll 1$ bubble nucleation is mainly driven by quantum fluctuations.
For $T\lesssim T_{c}$ the two phases have almost degenerate $\mathcal{F}$ and the \emph{thin wall} regime applies. Moreover, it can be seen that quantum bubbles are highly disfavoured so that tunneling is driven by thermal fluctuations. Nevertheless, as was shown in Ref.~\cite{Creminelli:2001th}, the tunneling rate is too slow for the transition to be completed in this regime. As the plasma cools to $T\ll T_c$ the properties of the critical bubble become very different. Firstly, the energy difference between the two phases increases and \textit{thick wall} bubbles become favoured. Secondly, especially in the regime of long supercooling which we are interested in, thermal bubbles production is either disfavoured or comparable to quantum induced one, as we checked also by numerically solving the bounce equations for the two cases. Thus in the following we will focus on quantum tunneling in the thick wall regime.

From Fig.~\ref{potential} we see that
 for lower and lower temperatures, the thermal barrier becomes smaller and smaller. In this regime it is not obvious where the field prefers to tunnel: whether it nucleates close to the minimum $\langle\mu\rangle\sim$ TeV  or just beyond the barrier (that is, roughly, at $\mu\sim T$) depends on the fine details of the potential $V_{\rm eff}$.
The task of finding the tunneling point $\mu_t$ (see Fig.~\ref{potential}), as long as the bounce action $S_B$, can be tackled by means of a thick wall approximation for $S_{B}$~\cite{Nardini:2007me} that, for a canonically normalized field $\phi$, reads as
\begin{equation}\label{thickwall}
S_B\approx \frac{2\pi^2}{3} \min_{\phi_t} \frac{|\phi_t-\phi_{fv}|^4}{V(\phi_{fv})-V(\phi_t)}\,,
\end{equation}
where $\phi_{fv}$ is the position of the false vacuum and the minimization is over the tunneling point $\phi_t$. In our specific case, \eq{thickwall} gives
\begin{equation}\label{twdilaton}
S_B\approx 24 N^2 \left(\frac{c_1}{12}\right)^2 \min_{\mu_t}\frac{\left(\sqrt{c_3/c_1}T+\mu_t\right)^4}{-16\pi^2 c_2 T^4-{\lambda}(\mu_t)\mu_t^4}\,,
\end{equation}
where the prefactor comes from the non-canonical normalization of $\mu$.
We have checked that this formula is in good agreement with the numerical results obtained by solving the bounce equation in the limit $T\ll T_c$. 

Let us now work out the consequences of \eq{twdilaton} for the models described in Sec.~\ref{SIS}.
\begin{itemize}
\item{\bf Type \I Models:}  
In this case ${\lambda}(\mu)$ is given by \eq{logdep},  that can be conveniently rewritten as 
${\lambda}(\mu)=\beta_\lambda (\ln (\mu/\Lambda_c)-1/4)$.
In the limit $\lambda(\mu_t)\ll 1$, which is a consequence of $\beta_\lambda\ll 1$, \eq{twdilaton} can be  solved,  giving $\mu_t\simeq T\, \Lambda_c/T_c$ and
\footnote{We notice that the terms   proportional to $\sqrt{c_3/c_1}$ in   \eq{twdilaton}  
give subleading contributions for $\lambda(\mu_t)\ll 1$.}
\begin{equation}\label{Stype1}
S_B\approx \left(\frac{c_1}{12}\right)^2 \frac{24 N^2}{\beta_\lambda \ln\frac{T_c}{T}}\,.
\end{equation}
The bubble radius is roughly $T^{-1}$, so the prefactor $\mathcal{A}$ is estimated as $T^4$ and the condition for exiting inflation $\Gamma>H^4$ gives
\begin{equation}\label{exitlight}
\left(\frac{c_1}{12}\right)^2\frac{24 N^2}{\beta_\lambda \ln\frac{T_c}{T}}\lesssim 4 \ln\frac{T M_P}{T_c^2}.
\end{equation}
The temperature at which the above condition is satisfied gives the exit temperature $T_e$, which can be conveniently traded for the number of efolds of inflation as follows
\begin{equation}\label{Ne}
N_e = \ln \frac{T_c}{T_{e}} \approx \frac{1}{2}\left( 1-\sqrt{1-\frac{24N^2(c_1/12)^2}{\beta_\lambda \ln^2\frac{M_P}{T_c}}} \right)\ln\frac{M_P}{T_c}\lesssim 17\,.
\end{equation}
When the argument of the square root in \eq{Ne} is negative, there is no solution to \eq{exitlight}, meaning that
the universe gets trapped in the deconfined phase.  This occurs whenever
\be
N\gtrsim 1.2 \left(\frac{12}{c_1}\right)\left(\frac{\beta_\lambda}{1/35}\right)^{1/2}\,,
\label{exitlicht}
\ee
where we normalized $\beta_\lambda$ according to the discussion after \eq{hierarchy}. 
We see that exiting supercooling in these type of models  is in tension with having a small $\beta_\lambda$,
 which is required to generate the hierarchy between $\lc$ and $M_P$.
 We could imagine more favorable situations with   $\beta_\lambda$  not  being constant,  for example
being small at the UV but  rapidly growing towards $\mu\lesssim \lc$ (the range of $\mu$ which is important for discussing supercooling).  This would allow 
 to exit supercooling for larger values of $N$, as appreciated in Fig.~\ref{exit} where we  show   $T_e$ for different values of  $N$ and $\beta_\lambda$.
 Nevertheless, having small $T_e$ (a long period of supercooling) 
 seems to require precise values of  $N$ and $\beta_\lambda$, which makes this quite  unlikely.

\begin{figure}[t]
	\center
	\includegraphics[scale=.18]{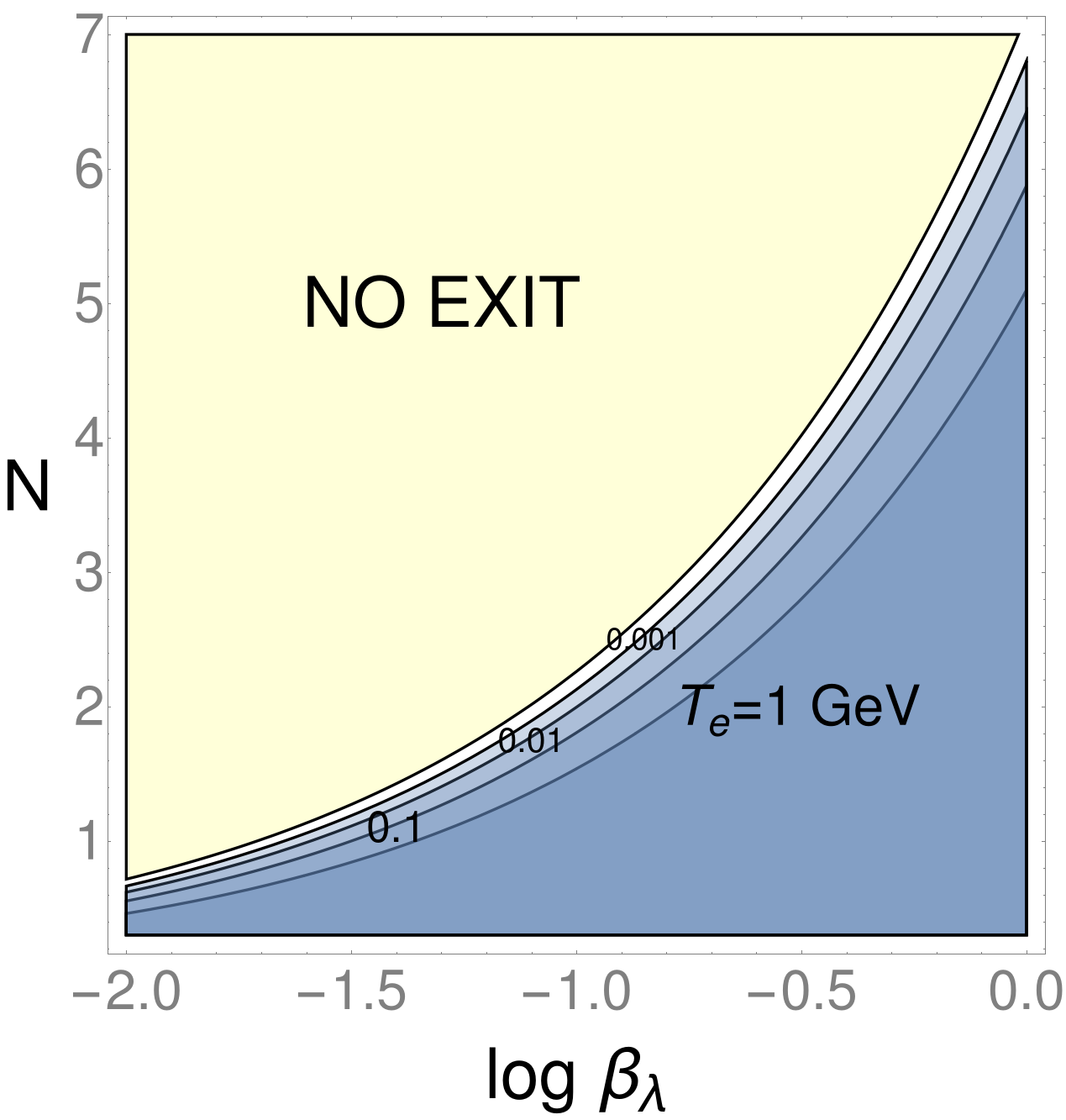}
	\caption{\it Lines of fixed exit temperature ($T_{e}=1,0.1,0.01,0.001$ GeV from below to above) for type \I models, and no exit region \eq{exitlicht}.}
	
	\label{exit}
\end{figure}

\item{\bf Type \II Models:} 
These models  have a larger $\beta_\lambda\sim 1$ (for $\mu\sim$ TeV), which leads to faster tunneling rates. Using \eq{logdep2} with $\lambda_0\lesssim \beta_\lambda$, we find that exit from supercooling  roughly happens for
$
N\lesssim 1-2\left( {12}/{c_1} \right) \beta_\lambda^{{1}/{2}}.
$
When the parameters allow this, inflation ends for $T_e$ not far from $T_c$.
It is difficult to achieve small values of $T_e$ since  $\lambda(\mu)$  becomes almost constant  for small $\mu$,
as it is shown in Fig.~\ref{lambdaplot}.
 In other words,  we find once more that long periods of supercooling are not expected.
\end{itemize}


The above conclusions can change drastically when  $T$  drops close to  $\lqcddec$.
As we will see  in the next section, at that scale QCD confines and can drive the supercooling era to an end.


 
\subsection{Impact of  QCD  at low $T$}\label{sec:qcd}

If the universe is trapped in the CFT deconfined phase, the temperature will drop exponentially till 
reaching the QCD strong scale, where  QCD effects could potentially become important. 
 The relevance for supercooling of QCD condensates was first discussed in Ref.~\cite{Witten:1980ez} in the context of the Coleman-Weinberg model (for recent studies see~\cite{Iso:2017uuu, Arunasalam:2017ajm, Bai:2018vik, Marzo:2018nov}). Nevertheless, there is an important difference with the models discussed here.
The Coleman-Weinberg is a weakly-coupled model where the scalar (which can be considered a dilaton with a potential as \eq{potdilaton}) is an elementary particle.  For this reason a huge fine-tuning is required to put to zero the mass term that, on general grounds, could be present. The models discussed here however are strongly-coupled theories in which the dilaton is composite  and therefore the potential \eq{potdilaton} arises naturally without fine-tuning.

The first analysis of the impact of QCD  in models of composite dilatons was done in Ref.~\cite{vonHarling:2017yew}. The authors study the contributions of the QCD vacuum energy  to the dilaton potential. This is  estimated to be
\begin{equation}\label{vqcd}
\Delta V_{\rm eff}(\mu)=V_{\rm QCD}(\mu)\simeq -c_4\frac{N_c^2}{16\pi^2}\Lambda_{\rm QCD}^4(\mu)\,,
\end{equation}
where $c_4\approx 7$  can be extracted from QCD data (taking $N_c=3$ and $\lc=330$ MeV), 
and  $\lqcd(\mu)$ is the QCD strong scale as a function of $\mu$.
This   dependence   on the value of $\mu$ arises,  as explained in Sec.~\ref{sec:dec},
from the  CFT states charged under QCD that affect the running of $g_s$  from high-energies down to  $Q\sim\mu$. 
Indeed, using  \eq{gsrun}, it can be shown that
\bea
 \Lambda_{\rm QCD}(\mu)&\sim&\mu^n\  \ \ \  \ \ \ \text{for}\  \ \mu>\lqcddec\,,\nonumber\\
 \Lambda_{\rm QCD}(\mu)&\sim&\lqcddec \  \ \ \text{for}\ \ \mu<\lqcddec\,,
 \eea
where  $n$ is given in  \eq{n}.
For $n<1$, \eq{vqcd}  can  become larger than $\lambda(\mu)\mu^4$ at small $\mu$.
For example, at $\mu\sim  \lqcddec$, where $\Delta V_{\rm eff}/V_{\rm eff}$ is maximized,  this happens for
\be
c_4 N_c^2>{\lambda}(\lqcddec)N^2\,.
\label{qcdabove}
\ee
If this is the case,  \eq{vqcd} dominates the dilaton potential  and the tunneling rate can be estimated as 
\footnote{This is  is obtained by solving \eq{twdilaton} with  \eq{vqcd}, taking into account that in the deconfined phase there is also the  contribution  $\Delta\mathcal{F}\simeq -c_4 (N_c^2/16\pi^2)(\lqcddec)^4$. The tunneling point $\mu_t$ is of order $\lqcddec$ and the bubble size  $\sim1/\lqcddec$.} 
\begin{equation}
S_B(T\sim \lqcddec)\approx \frac{24 N^4 (c_1/12)^2}{c_4 N_c^2 n (1-n)^{\frac{1}{n}-1}}\,,
\label{sbqcd}
\end{equation}
and the condition for exiting supercooling at $T\sim\lqcddec$ being 
 $S_B\lesssim 4\log M_P \lqcddec/\lc^2$.
Since \eq{sbqcd} scales as $N^4/N_c^2$, exit from
supercooling is only possible for small values of $N$.
In particular, we find $N\lesssim 2-3$ for $N_{c}=3$ and $n\sim 0.1-0.5$.
For a more detailed exploration see however Ref.~\cite{vonHarling:2017yew}.

\subsubsection{Fate of the new strong sector for $\mu$   below $\lqcddec$}\label{sub:qcdnew}

\begin{figure}[t]
	\center
	\includegraphics[scale=.55]{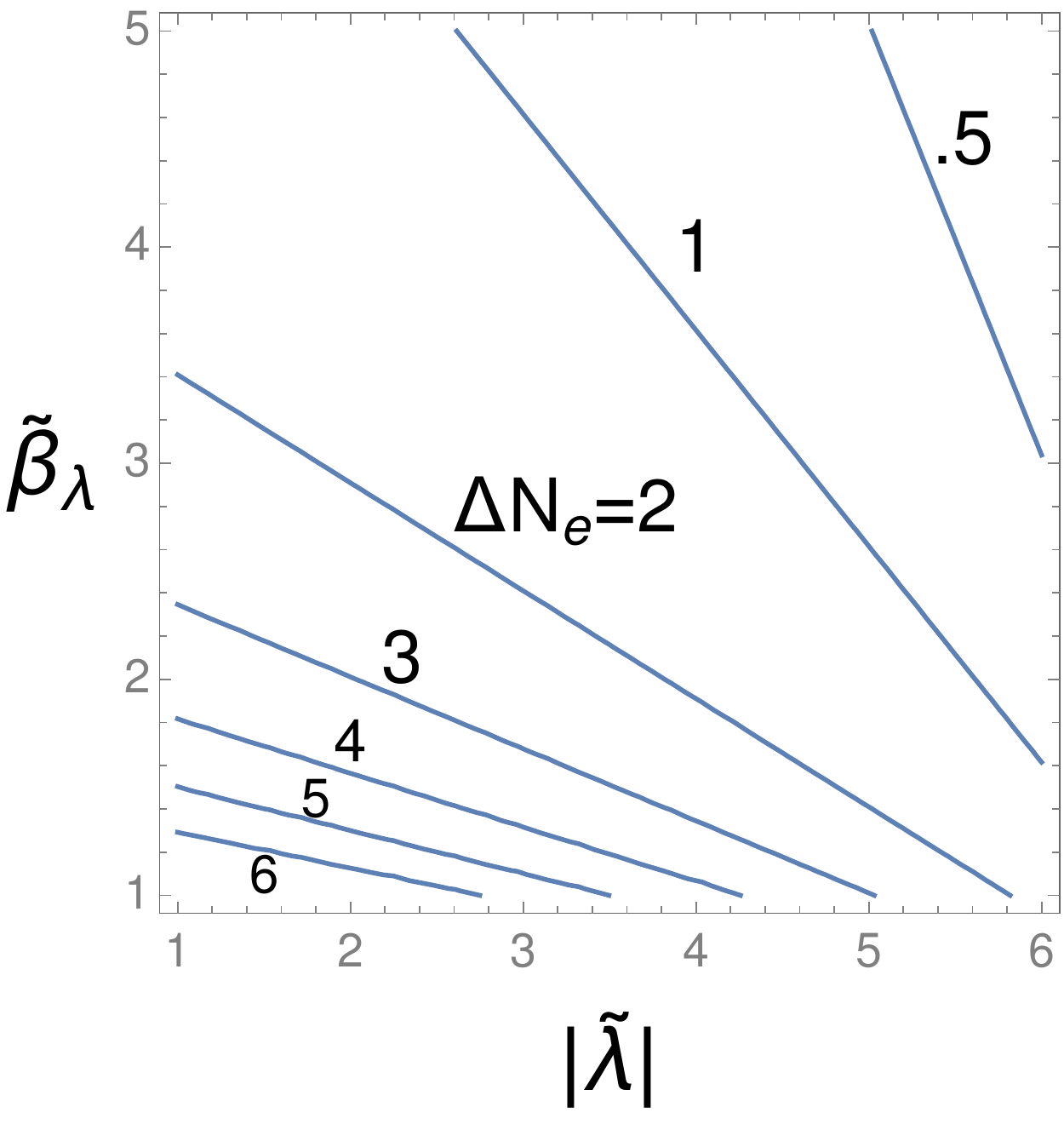}
	\caption{\it Contours of $\Delta N_e$, the number of efolds after the QCD phase transition, in the scenario described in Sec.~\ref{sub:qcdnew}. We take $N=7$ and $N_e=\ln(T_c/\lqcddec)\simeq 4.6$.}
	
	\label{efolds}
\end{figure}

As $T$ goes below $\lqcddec$, we expect the impact of QCD in the tunneling rate to be more dramatic,
as the CFT  can be  drastically affected.
Since some of  the CFT constituents are charged under color, 
we expect them to take mass at $\lqcddec$ and decouple at around this scale, analogously to hadrons in the SM below the GeV. We cannot really compute what is the fate of the CFT  at smaller scales, but it is not unconceivable that, after this drastic departure from near-conformality, the theory flows into another (quasi) CFT. An example of this behavior is provided by QCD in the conformal window, where the theory can remain conformal even after integrating out some flavors.
%
If this is the case, 
we can parametrize this effect as a drastic change of the quartic coupling 
${\lambda}\rightarrow {\lambda}+\Delta{\lambda}\equiv \tilde{\lambda}$ at $\lqcddec$, 
as well as a change in its beta-function, ${\beta_\lambda}\ll 1\rightarrow  {\tilde{\beta}_\lambda}\sim 1$.
Taking $\tilde{\lambda}$ and $\tilde{\beta}_\lambda$ as free parameters, the tunneling rate will be given by
\begin{equation}
\label{eq:actionqcd}
S_B\approx \frac{24 N^2(c_1/12)^2}{|\tilde{\lambda}|+\tilde{\beta}_\lambda \ln\frac{\lqcddec}{T}}\,,
\end{equation}
valid under the conditions that $\Delta{\lambda}<0$ and $\tilde{\beta}_\lambda>0$ (which are also the conditions that favor tunneling for $T<\lqcddec$).
Notice that, contrary to \eq{sbqcd}, the bounce action  is not enhanced by $N^2/N_c^2$,
which makes the supercooling exit much more probable. 
In Fig.~\ref{efolds} we plot the contours of fixed $\Delta N_e$, the efolds of inflation after the QCD phase transition is crossed, in the parameter space of $|\tilde{\lambda}|$ and $\tilde{\beta}_\lambda$, where we have taken $\lqcddec= 10^{-2}\, T_c$.

We  conclude that under  reasonable assumption on  the strongly-coupled CFT
 the universe could be trapped in the unconfined phase  till it reaches  $T\sim \lqcddec$.  
 At this temperature, or slightly below (see Fig.~\ref{efolds}), the tunneling rate to the CFT confined phase
 can be large enough due to  QCD, leading to the end of this long supercooling epoch.

\section{Cosmological Implications}
\label{sec:cosmology}

How would we know if the universe indeed experienced such a long supercooling epoch?   
The purpose of this section is to describe observational consequences of the associated long period of vacuum domination, which would provide evidence for this phenomenon (see also~\cite{Randall:2006py, Kaplan:2006yi, Nardini:2007me, Hassanain:2007js, Konstandin:2010cd, Konstandin:2011ds, Konstandin:2011dr, Bunk:2017fic, Dillon:2017ctw, Kobakhidze:2017mru, Bruggisser:2018mus, Bruggisser:2018mrt, Megias:2018sxv, Baldes:2018emh, Ellis:2018mja, Prokopec:2018tnq, Miura:2018dsy} for related analyses but where mainly only short eras of supercooling were considered).

Let us start with the universe at high-temperatures, reheated after a first epoch of inflation, responsible for the generation of the CMB anisotropies. As the temperature cools down and reaches $T_{c}\sim O(100)~\text{GeV}$, the universe gets trapped in the CFT deconfined phase, and a second inflationary epoch begins with Hubble rate given by~\eq{hubble}.
Following  the discussion in Sec.~\ref{sec:qcd}, we will mainly consider that the exit from supercooling occurs at or slightly after QCD confinement, i.e.~after $\sim 7-\ln(\lqcddec/330~\text{MeV})$ efolds.\footnote{The duration of supercooling is in principle not strongly constrained by CMB anisotropies, because density and/or isocurvature fluctuations are not abundantly produced~\cite{Turner:1992tz}.}  At the end of the supercooling era, there is an injection of entropy due to the first-order phase transition. While reheating (RH) in this case can be in general a complicated process, the natural expectation is that most of the energy available in the deconfined phase is transferred to radiation in the confined phase. This leads to
\begin{equation}
\label{eq:trh}
\mathcal{F}_{\text{dec}}(T_{c})\simeq \frac{\pi^{2}}{90}g_{*}T_{\text{RH}}^{4} \quad \Rightarrow\quad  T_{\text{RH}}\simeq \left(\frac{45}{4}\right)^{1/4}\frac{\sqrt{N}}{g_{*}^{1/4}}T_{c}\,.
\end{equation}
where we have used \eq{fdec} with $c_{2}=\pi^{2}/8$. Here $g_{*}\gtrsim 100$ is the number of relativistic species after reheating. Thus the entropy injection after supercooling is given by
\begin{equation}
\label{eq:entropy}
\frac{s_{\text{RH}}}{s_{e}}\simeq \left(\frac{4}{45}\right)^{1/4}\frac{g_{*}^{1/4}}{\sqrt{N}}\left(\frac{T_{c}}{T_{e}}\right)^{3}\,,
\end{equation}
where $s_{\text{RH}}$ and $s_{e}$ are respectively the entropy density at reheating and $T_e$; the latter can be read from \eq{freeooe} evaluated at $T_{e}$.
This release of entropy strongly dilutes the abundances of constituents which exist in the deconfined phase. This is in particular the case for particles heavier than the TeV scale as well as for baryons, as we will discuss in  Sec.~\ref{sub:dm}  and  Sec.~\ref{sub:baryons} respectively. The same conclusion applies to quanta of a QCD axion field, if the latter oscillates during supercooling (see Sec.~\ref{sub:misalignment}). However, the relevance of axionic topological defects is generically enhanced in the supercooled universe, as we will discuss in Sec.~\ref{sub:topological}. 

At the end of the first order phase transition, reheating occurs via violent processes, which source gravitational radiation. While this implication has already been thoroughly discussed in the literature (see refs. above), our long epoch of inflation can lead to a maximal signal, dominantly due to the collision of bubbles of the confined phase, as we will discuss in Sec.~\ref{sec:gws}.

\subsection{Dark Matter abundance}
\label{sub:dm}

Let us start with the implications for dark matter (DM) particles. In the strongly-interacting scenarios considered above, we can consider two classes of dark matter candidates (apart from the QCD axion that will be discussed in the next section):
those which are heavy in both CFT phases, or, alternatively, those which are massive in the confined phase but massless in the deconfined phase, i.e.,  $m\propto \mu$. In both cases, supercooling affects the relic abundance only if the dark matter is not repopulated after reheating. If a generic DM particle $\chi$ has mass $m_{\chi}$ and annihilation cross section $\sigma_{ann}$ in the confined phase, its thermal production after supercooling is subdominant whenever $T_{\text{RH}}\lesssim T_{\text{fo}}$, where $T_{\text{fo}}$ is the freeze-out temperature of $\chi$. For $\sigma_{ann} m_{\chi}\geq 10^{-16}~\text{GeV}^{-1}$, one finds $T_{\text{fo}}\sim O(0.1)~m_{\chi}$. Therefore, according to \eq{eq:trh}, thermal production after supercooling can be neglected when $m_{\chi}\gtrsim \text{TeV}$. In what follows we will focus on this case.\footnote{Non-thermal production after supercooling can also be relevant in principle. Indeed, in a first-order phase transition with runaway bubbles, the possibility exists to abundantly produce heavy dark matter particles during the bubble collisions~\cite{Konstandin:2011ds, Falkowski:2012fb, Katz:2016adq}. This occurs only when collisions are approximately elastic. However, in our case the two minima of the dilaton potential are highly non-degenerate, which implies that bubble collisions are likely quite inelastic. This remains true even if the field tunnels to a value of the potential which is close to the false minimum, since it rolls down the true minimum before the bubbles have time to meet. Therefore non-thermal production of heavy dark matter from bubble collisions is suppressed in our scenarios.}

Let us consider first the dilution of  DM candidates that are massive  in both phases,  $m_{\chi}^{\text{dec}}\simeq m_{\chi}$, and whose thermal freeze-out temperature is above $T_{c}$.  Their relic abundance today can be computed starting from the number density-to-entropy ratio $Y_{\chi}\equiv n_{\chi}/s$. After freeze-out, $Y_{\chi}$ is constant, except at the phase transition where the entropy changes according to \eq{eq:entropy}. Therefore, the relic abundance of $\chi$ is given by
\begin{align}
\label{eq:relicdm}
\nonumber \Omega_{\chi} h^{2} &\simeq 0.1~ \frac{\sqrt{N}}{g_{*}^{1/4}}\left(\frac{T_{e}}{T_{\text{c}}}\right)^{3} \left(\frac{m_{\chi}^{\text{dec}}/T_{\text{fo}}}{20}\right)\left(\frac{10^{-8}~\text{GeV}^{-2}}{\sigma_{ann}}\right)\\
&\simeq 0.1~  \frac{\sqrt{N}}{g_{*}^{1/4}}\left(\frac{T_{e}}{157~\text{MeV}}\right)^{3} \left(\frac{m_{\chi}^{\text{dec}}/T_{\text{fo}}}{20}\right)\left(\frac{10^{-18}~\text{GeV}^{-2}}{\sigma_{ann}}\right).
\end{align}
In the second line of \eq{eq:relicdm} we have assumed $T_{c}\simeq 300~\text{GeV}$ and normalized $T_e$  to the SM QCD critical temperature. \eq{eq:relicdm} is simply the standard relic abundance of a WIMP-like particle  with a very small prefactor due to supercooling. If we assume $\sigma_{ann}\sim g_{\chi}^{4}/(m_{\chi}^{\text{dec}})^2$ and $m_{\chi}^{\text{dec}}\sim~\text{TeV}$, \eq{eq:relicdm} implies $g_{\chi}\lesssim 10^{-3}$   for $T_e\sim$ 100 MeV.\footnote{Heavy particles which interact with nuclei can be looked for at DM direct detection experiments. Interestingly, if $\sigma_{\text{ann}}\simeq \sigma_{\chi-\text{nucleon}}$, the region of parameter space suggested by \eq{eq:relicdm} is currently probed by XENON1T~\cite{Aprile:2018dbl} and will be further investigated by LZ~\cite{Mount:2017qzi}.} 
If we assume instead $g_{\chi}\sim O(1)$, \eq{eq:relicdm} matches the observed  DM  abundance for $m_{\chi}^{\text{dec}}\simeq 10^{9}$~GeV. Thus in the supercooled universe \emph{wimpzillas} do not need to be very weakly coupled in order not to overproduce dark matter. 

Let us now move to species which are massless in the deconfined phase $m_{\chi}^{\text{dec}}=0$ and massive  in the confined phase,  $m_{\chi}\sim \mu$. We only demand these states to be stable.
Their relic abundance  can be computed again from  $Y_{\chi}$, which in the deconfined phase is given by $ Y_{\chi}\sim g_{\chi} T^3/s$,  with  the entropy $s$ given in  \eq{freeooe}.
The number density of $\chi$ today can be thus simply obtained by dividing $Y_{\chi}$ by \eq{eq:entropy} and multiplying  by the entropy density today. The relic abundance of $\chi$ is thus given by
\begin{equation}
\label{eq:relicmassless}
\Omega_{\chi}h^{2}\simeq 0.5\, \frac{g_{\chi}\kappa_{\text{DM}}}{N^{3/2}}\left(\frac{100}{g_{*}}\right)^{1/4}\left(\frac{m_{\chi}}{\text{TeV}}\right)\left(\frac{T_{e}}{157~\text{MeV}}\right)^{3},
\end{equation}
where $\kappa_{\text{DM}}\lesssim 1$ is an efficiency prefactor which takes into account the possibility that $\chi$ is further diluted as its mass switches on during the phase transition. If the DM is a composite state in the confined phase, \eq{eq:relicmassless} should include a further $O(1)$ prefactor to take into account the decrease of degrees of freedom in DM from the deconfined to the confined phase. From \eq{eq:relicmassless}  we can conclude that a TeV-scale particle  can reproduce the observed dark matter abundance if supercooling lasts approximately until the temperature of the SM QCD phase transition (see~\cite{Hambye:2018qjv} for similar conclusions). 

\subsection{Coincidence between baryon and dark matter abundances}
\label{sub:baryons}

Similarly as dark matter, any baryon asymmetry generated at high temperature will be diluted by the inflationary expansion during supercooling. This opens the possibility to explain the smallness of $(n_B-n_{\bar B})/s$ by the dilution factor \eq{eq:entropy}.
Furthermore,  if both  dark matter   and baryons arise from  relativistic particles that were in  thermal equilibrium 
at some $T>T_c$ , their abundances  will be  comparable before supercooling and equally diluted at the end of the phase transition.
This could  then provide  an explanation  for the coincidence of DM and baryon abundances, with their energy density  ratio mainly  determined by  the ratio of their masses.

As an illustrative  example of the above idea, let us consider the case in which a (B-L) asymmetry is generated at temperatures above $T_{c}$,
from the out-of-equilibrium decay of a very heavy state $\Psi$ (generic in UV completions which include gravity).
For $T>M_{\Psi}$, $\Psi$ is in thermal equilibrium, thus its number density is comparable to the rest of the thermal bath, i.e. $n_{\Psi}\propto g_{\Psi} T^{3}$. Once $T\lesssim M_{\Psi}$,  the inverse decays are Boltzmann-suppressed and a  baryon asymmetry can be generated if the out-of-equilibrium condition $\Gamma_{\Psi\to {\text{B-L}}}\lesssim H(T\sim M_{\Psi})$ is fulfilled. 
 If this is the case, we expect $n_{\text{B-L}}\simeq n_{\text{B}}\sim \epsilon_{\text{CP}}n_{\Psi}$, where $\epsilon_{\text{CP}}$ is the CP asymmetry in the decay of $\Psi$ to (B-L)-charged particles. After supercooling, the baryon number is diluted in the same way as in the case of DM.
Following the same steps which lead to \eq{eq:relicmassless}, we can estimate  the baryon-to-entropy ratio to be
\begin{align}
\label{eq:baryons}
\frac{n_{\text{B}}}{s}\sim 0.01~\frac{\epsilon_{CP}}{N^{3/2}}\left(\frac{100}{g_{*}}\right)^{1/4}\left(\frac{T_{e}}{T_{c}}\right)^{3}.
\end{align}
If $\epsilon_{CP}\sim O(1)$ and $N\gtrsim 5$, the observed baryon asymmetry can be obtained for $T_{e}\sim 1~\text{GeV}$.
As the origin of \eq{eq:baryons} is similar to DM in \eq{eq:relicmassless}, we can easily estimate the ratio of  baryon  to  DM energy density
 to be
\begin{equation}
\label{eq:coincidence}
\frac{\Omega_{B}}{\Omega_\chi}\simeq \frac{\epsilon_{CP}}{\kappa_{\text{DM}}}\frac{m_{p}}{m_{\chi}}\,,
\end{equation}
where $m_p$ is the proton mass and we have also included a possible dilution factor for $\chi$, as in \eq{eq:relicmassless}. 
It is quite interesting that the ratio of  baryon and DM relic abundances is  predicted to be very simply related to  the ratio of their masses and of the efficiencies of CP violation and DM number changing processes at the end of the phase transition. 
In order to obtain the observed value $\Omega_{\chi}\simeq 5\Omega_{B}$, \eq{eq:coincidence} leaves two options:
\begin{itemize}
\item[$\mathbf{1.}$]{$m_{\chi}\sim 5~\text{GeV}$ with $\epsilon_{CP}\sim \kappa_{\text{DM}}\sim O(1)$. However, in this case one needs to make sure that the DM is not copiously produced at reheating and that its supercooling abundance is not overcome by later thermal or sub-thermal production. As explained above, this forces us to consider $T_{\text{RH}}\ll m_{\chi}$, which might be  challenging to achieve.}
\item[$\mathbf{2.}$]{$m_{\chi}\gtrsim \text{TeV}$ with $\kappa_{\text{DM}}/\epsilon_{CP}\sim 10^{-3}$. This implies that  the DM abundance has to be further diluted during the phase transition. This is not inconceivable: in the case of the electroweak phase transition it is indeed known that at the bubble walls, particles can undergo transition radiation and lose energy via the emission of massive particles~\cite{Bodeker:2017cim}. While we do expect the same phenomenon to occur in the deconfined-confined transition, we are not currently  aware of quantitative estimates.}
\end{itemize}
 
\noindent In the light of the arguments presented in this subsection, it would be extremely interesting to understand whether a low reheating temperature and/or small $\kappa_{DM}$ can be obtained in the supercooling setup. We leave this task for future work.

\subsection{The QCD axion}
\label{sec:axion}

Let us now discuss the implications of supercooling for the QCD axion. In the deconfined phase SM quarks can be either massless or massive, depending on whether the Higgs VEV (or that of the  equivalent operator ${\cal O}_H$) is zero or not in this phase. 
On one hand, in models of type \Ico  all VEVs are zero in the deconfined phase and quarks are then massless. In this  case  the axion  mass is quite small  in the deconfined phase\footnote{Even for $\langle H\rangle=0$, the axion mass is  not zero since  there is still a very small Higgs-mediated contribution.} and supercooling does not alter the relic abundance from the misalignment mechanism~(see~\cite{Servant:2014bla} for previous studies).
On the other hand, in models of type \IIco we have $\langle {\cal O}_H\rangle\sim \lssb\sim$ TeV and the SM quarks are instead  massive in the deconfined phase. Axion oscillations can then take place during the supercooling inflationary expansion as well as $3H\lesssim m_{a}^{\text{dec}}$, where $m_{a}^{\text{dec}}$ is the axion mass in the deconfined phase. 
 In this case the axion field can rapidly relax to the minimum  and its relic abundance from the misalignment mechanism can be diluted, as we will explain in detail below.
  Axionic topological defects can instead be affected by supercooling in both type of models, as  it will be discussed at the end of this section.

The standard axion cosmology begins when the PQ symmetry is broken at $T\sim F_a$, where $F_{a}$ is the axion decay constant. Below this temperature, the axion dynamics is dictated by the Klein-Gordon equation
\begin{equation} 
\label{eq:kleingordon}
\ddot{a}+3H\dot{a}+V'(a, T)=0\,.
\end{equation}
The potential $V$ is periodic and negligible at high temperatures, while it rises very rapidly around the QCD phase transition:
\begin{equation}
\label{eq:potential}
V(a, T)\simeq \frac{m_a^2(T)F_{a}^{2}}{N_{\text{DW}}^{2}}\left[1-\cos\left(\frac{N_{\text{DW}} a}{F_a}\right)\right],
\end{equation}
with $m_a^{2}(T)=m_{a}^{2}(0)(T_{\text{QCD}}/T)^{n}$ and $n\approx 8$~(see~\cite{Gross:1980br} for an analytic determination and~\cite{Borsanyi:2016ksw} for lattice confirmation). Below $T_{\text{QCD}}\simeq 157~\text{MeV}$, the axion mass is given by
\begin{equation}
\label{eq:zeromass}
m_{a}^{2}(0)=\frac{m_{u}m_{d}}{(m_{u}+m_{d})^{2}}\frac{m_{\pi}^{2}f_{\pi}^{2}}{F_a^{2}}\simeq \frac{m_{u}m_{d}}{(m_{u}+m_{d})}\frac{\Lambda^{3}_{\text{QCD}}}{F_a^{2}}\,.
\end{equation}
According to \eq{eq:kleingordon}, axion oscillations start at $T=T_{\text{osc}}$ determined by $3 H(T_{\text{osc}})\simeq m_a(T_{\text{osc}})$, i.e.~at $T_{\text{osc}}\sim$~GeV for standard axion parameters. Below $T_{\text{osc}}$ axion quanta behave as DM, according to the misalignment mechanism. 
The QCD axion relic abundance depends on whether the PQ symmetry is broken during or after cosmological inflation. In the latter case, the initial average value of the axion field is $\theta_{i}\equiv a_{i}/F_{a}=\pi/\sqrt{3}$ and axion oscillations generate the observed dark matter abundance for $F_{a}\simeq 10^{11}~\text{GeV}$. If instead the PQ symmetry is broken during inflation, larger values of $F_{a}$ are phenomenologically viable if $\theta_{i}$ is appropriately tuned.

\subsubsection{Supercooling and the misalignment mechanism}
\label{sub:misalignment}

Let us now investigate the effects of supercooling on the misalignment mechanism. We approximate \eq{eq:potential} with the quadratic expansion around the minimum, since the effect of anharmonicities is not especially relevant for our estimates. The value of the axion field at the end of the supercooling era depends on the axion mass \eq{eq:zeromass} with $v\rightarrow v^{\text{dec}}, \Lambda_{\text{QCD}}\rightarrow \Lambda_{\text{QCD}}^{\text{dec}}$, the Hubble scale~\eq{hubble}, and the number of efolds of supercooling.
In particular, \eq{eq:kleingordon} admits three regimes for the amplitude of the axion field:
\begin{align}
\label{eq:under}
\textbf{Underdamping:}\ \ 3H\ll 2m_{a}; &\quad \theta_{f}\sim \theta_{i}\left[\frac{m_{a}(T_{\text{osc}})}{m(T_{e})}\right]^{1/2}e^{-\frac{3}{2}N_{e}}\,,\\
\label{eq:critical}
\textbf{Critical damping:}\ \  3H\simeq 2m_{a}; &\quad \theta_{f}\sim \theta_{i}e^{-\frac{3}{2}N_{e}}(1+\frac{3}{2}N_{e})\,,\\
\label{eq:over}
\textbf{Overdamping:}\ \ 3H\gg 2m_{a}; &\quad \theta_{f}\sim \theta_{i}e^{-\frac{m_{a}^{2}}{3 H^{2}}N_{e}}.
\end{align}
The underdamped case corresponds to the well-known dark matter behavior $\rho \sim a^{-3}$, with a prefactor due to the temperature dependence of $m_{a}$.
The condition $3H\lesssim m_{a}(0)$  provides an upper bound on $F_a$: 
\begin{equation}
\label{eq:fmax}
F_{a}\lesssim \frac{10^{11}~\text{GeV}}{N}~ \left(\frac{1~\text{TeV}}{m_{\text{dil}}}\right) \left(\frac{1~\text{TeV}}{\Lambda_{c}}\right)\left(\frac{\Lambda^{\rm dec}_{\text{QCD}}}{0.3\, \text{GeV}}\right)^{3/2}\left(\frac{v_{\rm dec}}{\text{1\, TeV}}\right)^{1/2}\,.
\end{equation}
The implication of \eq{eq:fmax} is as follows: On one hand, if $F_a$ is smaller than \eq{eq:fmax}, then the axion oscillates during the supercooling epoch and relaxes to the minimum $a/F_{a}=0$ if $T_{e}\lesssim \Lambda^{\rm dec}_{\text{QCD}}$. On the other hand, if $F_a$ is much larger than \eq{eq:fmax}, the axion is slowly-rolling along its potential: according to \eq{eq:over}, it relaxes to the minimum only for $N_{e}\gtrsim 3H^{2}/m_a^{2}$.

\begin{figure}[t]
  \centering
 \includegraphics[width=1\textwidth]{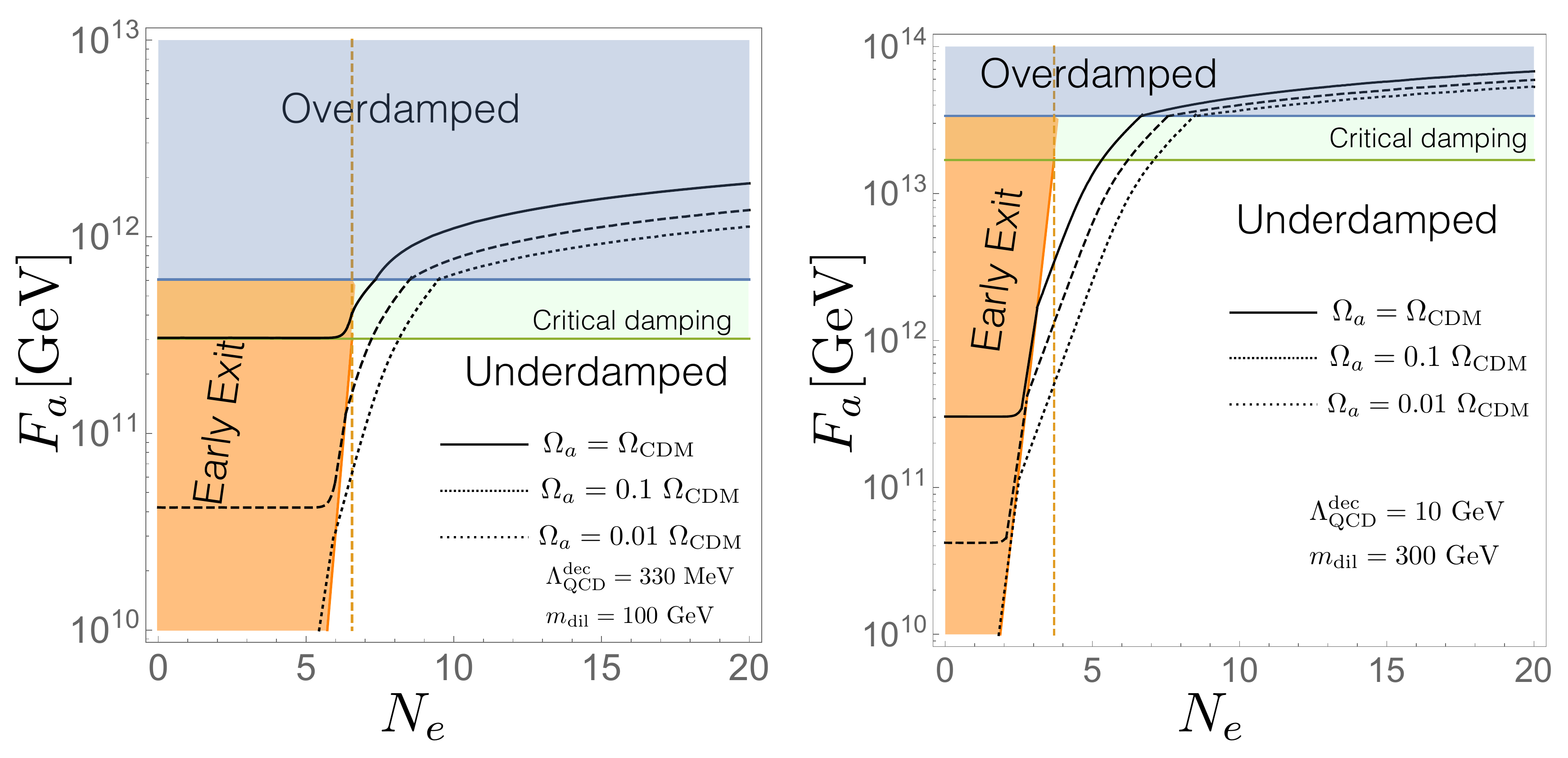}
 \caption{\it Constraints on $F_a$ and $N_{e}$ from DM overproduction. We have taken $\Lambda^{\rm dec}_{\rm QCD}=0.33~(10)~\rm{GeV}$ in the left (right) plot, as well as $N=5$. In the blue region we have $3H>2m_a$, in the green  region $3H\leq 2m_a$, and  in the white region $3H\leq m_a$.  In the orange region the supercooling era ends before the condition $3H=m_a$ is satisfied. Contours of $\Omega_{a}=1,\, 0.1,\, 0.01\, \Omega_{\text{CDM}}$ are respectively shown as thick, dashed and dotted black lines respectively. The vertical orange dashed line shows the number of efolds corresponding to the critical QCD temperature during supercooling.}
 \label{fig:nspace}
\end{figure}

The observationally allowed value of $F_{a}$ depend on the duration of supercooling. In Fig.~\ref{fig:nspace} we provide two examples of constraints on $F_a$ and $N_{e}$ from the overproduction of axion DM. We use $v^{\rm dec}\simeq 1$ TeV and fix $N=5, m_{\text{dil}}=100~(300)~\text{GeV}$\footnote{See~\cite{Blum:2014jca} for constraints on the dilaton mass from collider searches.}, $\Lambda^{\rm dec}_{\text{QCD}}=330~\text{MeV}$ ($10\ \text{GeV}$) in the left (right) plot in Fig.~\ref{fig:nspace}. 
The dilution of the axion field does not depend on the initial value of the axion but to fix ideas we have taken $a_i=(\pi/\sqrt{3})F_a$ as in the post-inflationary axion scenario. 
The dashed orange lines in the plots correspond to exiting at the QCD critical temperature, $T_{\text{QCD}}^{\text{dec}}\sim  0.5~\Lambda^{\rm dec}_{\text{QCD}}$. Assuming that the exit from supercooling occurs after the QCD phase transition, the upper bound on $F_a$ can be pushed slightly above $10^{12}~\text{GeV}$ for $\Lambda^{\rm dec}_{\text{QCD}}=330~\text{MeV}$. For $\Lambda^{\rm dec}_{\text{QCD}}=10~\text{GeV}$, $F_{a}$ slightly above $10^{13}~\text{GeV}$ is allowed.
The available parameter space is reduced as $\Lambda^{\rm dec}_{\text{QCD}}$ decreases and/or $m_{\text{dil}}$ increases, as dictated by \eq{eq:fmax}. In particular for $\Lambda^{\rm dec}_{\text{QCD}}\ll 300~\text{MeV}$ the axion field is essentially frozen during the supercooling epoch and the standard computation of the axion relic abundance is not significantly altered.
We also show in Fig.~\ref{fig:nspace} regions of parameter space where the QCD axion represents only a fraction of the DM. These are relevant for axion DM detection experiments (e.g.~ADMX~\cite{Du:2018uak}), which are sensitive to  $\sqrt{\Omega_{a}}$.

A few comments are in order. One can argue that the constraints on $F_a$ can also be relaxed in the standard case by demanding that the  PQ symmetry is broken during  inflation, as this allows to tune the initial condition $\theta_i$ to any desired value.
The mechanism presented here, however, provides a dynamical way to implement this initial condition, without appealing to any anthropic tuning.\footnote{See~\cite{Graham:2018jyp, Guth:2018hsa} for an alternative possibility in the framework of low-scale models of eternal inflation.}
Furthermore, our setup can still preserve the predictivity of the standard post-inflationary axion scenario. Indeed, the inflationary expansion during supercooling is too short to homogenize the axion field across the Hubble volume today, thus in our case one still has to average over randomly distributed axion values.


\subsubsection{Supercooling and axionic topological defects}
\label{sub:topological}

In the previous subsection we have focused on the relic abundance from the misalignment mechanism. However, further contributions to the axion energy density can arise from topological defects~(see~\cite{Vilenkin:2000jqa} for an excellent introduction). We now study how the latter are affected by supercooling. The standard lore is that inflationary expansion strongly dilutes pre-existing topological defects. However, we will argue here that if the PQ symmetry is broken after cosmological inflation, supercooling generically enhances the relevance of axionic strings and domain walls. 

The history of axionic topological defects starts at $T\sim F_a$, when the Peccei-Quinn symmetry is broken and string-like defects form. They evolve until $T_{c}$ in a radiation dominated background, according to the so-called \emph{scaling behavior}: at any given cosmological time there is approximately one horizon-length string per Hubble volume~(for a recent analysis, see~\cite{Klaer:2017ond, Gorghetto:2018myk, Kawasaki:2018bzv}). This fast dilution is partially due to the emission of massless axion quanta from $T\sim F_a$ to $T_{\text{osc}}$. The relic number density of such axions is dominated by particles emitted at $T_{\text{osc}}$. 

In the standard cosmological history, the axion potential \eq{eq:potential} switches on at the QCD phase transition and domain walls form, which are attached to the strings. Such walls pull the strings together and thus destroy the network if $N_{\text{DW}}=1$~(see~\cite{Kawasaki:2014sqa, Klaer:2017ond}). In the supercooling scenario strings are pushed beyond the cosmological horizon by the inflationary expansion.
At the QCD phase transition during supercooling, strings are not in causal contact and the network cannot annihilate, since two strings cannot feel the attractive force of the wall stretching between them. Therefore, the network survives the first QCD phase transition.

After reheating, as $T$ drops in the ordinary expanding universe, the string network re-enters the comoving Hubble sphere at the temperature $T_{\star}\simeq T_{e}$. There are now two possible scenarios depending on $T_{\star}$:
\begin{itemize}
\item[$\mathbf{1.}$]{$T_{\star}\gtrsim\text{GeV}$: the standard cosmology of topological defects and their contribution to axion dark matter is not significantly affected.}
\item[$\mathbf{2}$]{$T_{\star}\lesssim\text{GeV}$: in this case strings re-enter the horizon when $3H<m_{a}$. Therefore, the string-wall network is immediately formed and rapidly decays. The contribution to the axion dark matter abundance is given by (see~Appendix \ref{app:topological} for a derivation)
\begin{equation}
\label{eq:abundancenw}
\Omega_{\text{network}}h^{2}\simeq 0.03 \left(\frac{61.75}{g_{*}(T_{\star})}\right)^{1/4}\left(\frac{F_{a}}{10^{10}~\text{GeV}}\right)\left(\frac{100~\text{MeV}}{T_{\star}}\right).
\end{equation}
Constraints on $F_{a}$ and $T_{\star}$ from dark matter overproduction are shown in Fig.~\ref{fig:walls}. We are neglecting the contribution coming from the misalignment mechanism. This is justified for the relevant range of $F_a$ and $N_e$  (see Fig.~\ref{fig:nspace}). The effect of supercooling is   essentially to delay the collapse of the string-wall network, which leads to stronger bounds on $F_{a}$. \emph{En passant}, let us notice that such a delayed annihilation of axionic topological defects could lead to the formation of Primordial Black Holes~\cite{Ferrer:2018uiu}. 

\begin{figure}[t]
    \centering
    \includegraphics[width=0.6\textwidth]{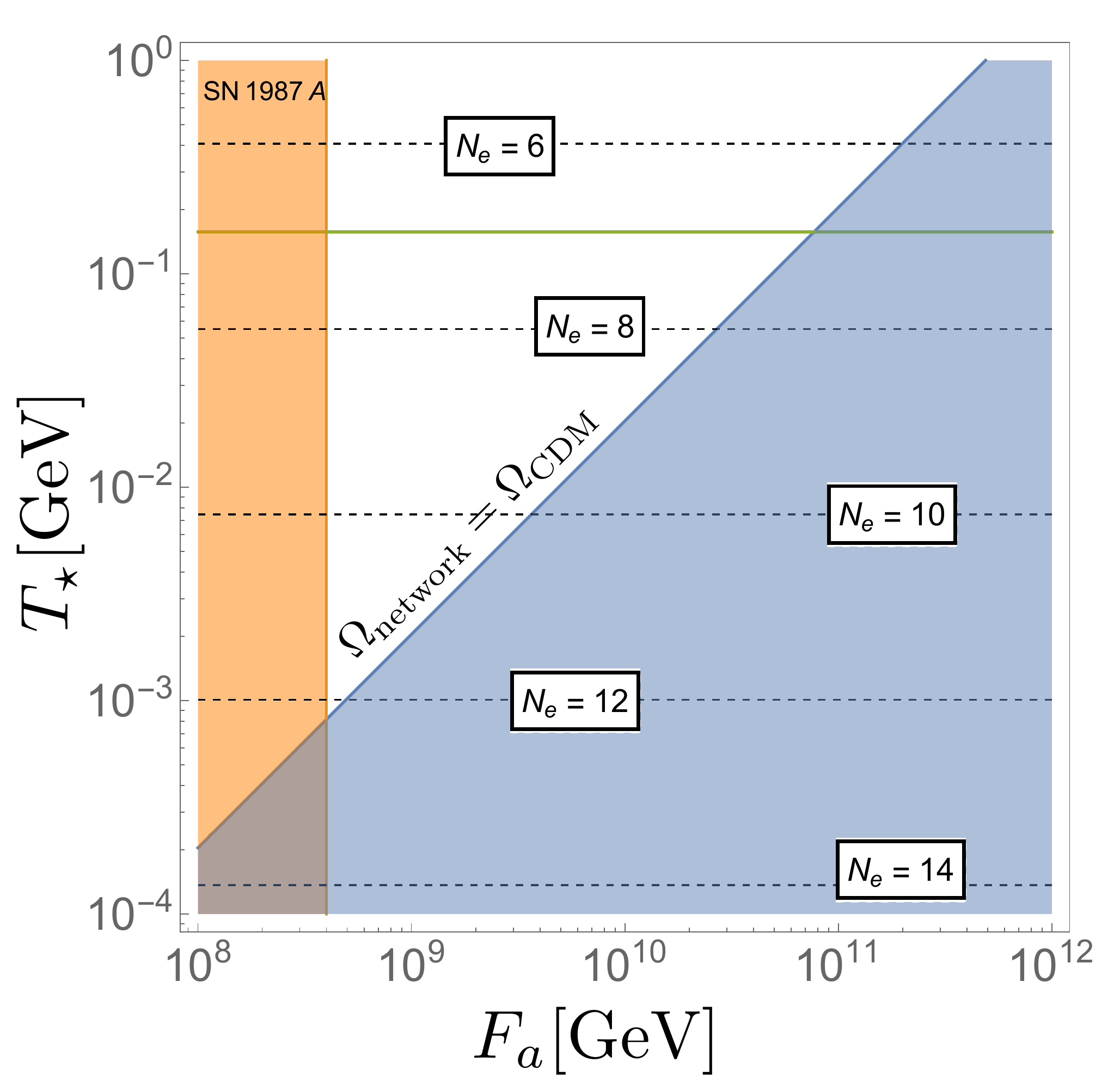}
	\caption{\it Constraints on $F_{a}$ and $T_{\star}$ from DM overproduction from the late decay of the axionic string-wall network. The solid blue line is obtained imposing $\Omega_{\text{network}}=\Omega_{\text{CDM}}$, according to~\eq{eq:abundancenw}. The orange  region is forbidden by astrophysical observations. The solid green line represent the critical temperature of the ordinary  QCD phase transition after reheating. The dashed lines show the corresponding number of efolds of supercooling.}
    \label{fig:walls}
\end{figure}}

\end{itemize}

\subsection{Gravitational waves}
\label{sec:gws}

The exit from the supercooling era occurs via a strong first order phase transition and thus generates gravitational radiation. Differently from other scenarios discussed in the literature, we focus here on the case in which the phase transition takes place after a long inflationary phase. As we will discuss, this implies that the gravitational wave signal is dominantly due to the collision of bubbles of the confined phase. A further peculiarity of our setup is the possibility to obtain the maximal strength of the signal, because of a long duration of the phase transition.

The gravitational wave signal depends mainly on two parameters: the duration of the phase transition, usually denoted with $\beta^{-1}$, and the latent heat $\alpha$ released during the phase transition. The former can be computed starting from the time variation of the nucleation rate, i.e.~$\beta\equiv 1/\Gamma(d\Gamma/dT)$ (see e.g.~\cite{Apreda:2001us}) and for the case of supercooling is given by~\cite{Caprini:2015zlo}
\begin{align}
\label{eq:beta}
\frac{\beta}{H(T_{\text{RH}})}=\frac{H_{\text{sc}}}{H(T_{\text{RH}})}\left[T_{e}\left(\frac{dS_{B}}{dT}\right)_{T_{e}}-4\right]\,.
\end{align} 
Here we denote with $H_{\text{sc}}$ and $H(T_{\text{RH}})$ the Hubble rate during supercooling~\eq{hubble} and  at the reheating temperature respectively.
For fast reheating they are expected to be similar, $H_{\text{sc}}\simeq H(T_{\text{RH}})$.
The latent heat is instead given by~\cite{Caprini:2015zlo}
\begin{equation}
\label{eq:alpha}
\alpha=\frac{V_{\text{eff}}(\langle\mu\rangle)}{\rho_{\gamma}(T_{e})}\,.
\end{equation}
In our setup $\alpha\gg 1$, while $\beta/H(T_{\text{RH}})$ can vary between $O(1)$ and $O(100)$. As we will see below, the strength of the gravitational wave signal is maximized for $\alpha\gg 1$ and $\beta/H(T_{\text{RH}})\sim 1$.

Until recently, it was thought that if a first order phase transition occurs during vacuum domination, bubbles of true vacuum would reach a runaway regime (see e.g.~\cite{Caprini:2015zlo}). This means that bubbles are accelerated until they actually collide, such that most of the energy available in the phase transition is deposited in the accelerating walls. If this is the case and bubbles expand at speeds very close to luminal, then a naive estimate of the wall $\gamma$-factor at the time of collision is $\gamma_{\text{max}}\sim \beta^{-1}R_{0}^{-1}$, where $R_{0}\sim T_{e}^{-1}$ is the initial bubble radius at $T_{e}$. This gives 
\be
\gamma_{\text{max}}\sim \left(\frac{H(T_{RH})}{\beta}\right)\frac{M_{P}}{\lc}\frac{T_{e}}{\lc}\sim (10^{13}-10^{15})\frac{T_{e}}{\lc}\,,
\ee
for the values of $\beta/H(T_{RH})$ which we find in our models. The subsequent collisions of several bubbles generate a strong gravitational wave signal, whose spectrum is given by~\cite{Caprini:2015zlo}~\footnote{After publication of this work, we became aware of analytical studies of the gravitational wave signal from bubble collisions~\cite{Jinno:2016vai, Jinno:2017fby}, which also go beyond some of the approximations assumed in~\eq{eq:gwbubbles}. For runaway bubbles, such as the ones we consider here, the gravitational wave spectrum at small frequencies turns out to scale as $\sim \omega$, while the peak frequency is slightly smaller than \eqref{eq:freqbubbles}.}
\begin{align}
\label{eq:gwbubbles}
\nonumber \Omega_{\text{GW}, \circ}h^{2} &\simeq 1.67\cdot 10^{-5}\left(\frac{H(T_{\text{RH}})}{\beta}\right)^{2}\left(\frac{\kappa_{\phi} \alpha}{1+\alpha}\right)^{2}\left(\frac{100}{g_{*}}\right)^{1/3}\left(\frac{0.11 v_{w}^{3}}{0.42+v_{w}^{2}}\right)S_{\circ}(\omega),\\
S_{\circ}(\omega)&=\frac{3.8(\omega/\omega_{\circ})^{2.8}}{1+2.8(\omega/\omega_{\circ})^{3.8}},
\end{align}
where $v_{w}\simeq 1$ is the bubble wall velocity, $\kappa_{\phi}\simeq 1$ is the fraction of the latent heat which goes into kinetic energy of the bubble walls and $\omega$ is the frequency, whose value at the peak of the signal is given by
\begin{equation}
\label{eq:freqbubbles}
\omega_{\circ}=1.65\cdot 10^{-2}\text{mHz} \left(\frac{0.62}{1.8-0.1 v_{w}+v_{w}^{2}}\right)\left(\frac{\beta}{H(T_{\text{RH}})}\right)\left(\frac{T_{\text{RH}}}{100~\text{GeV}}\right)\left(\frac{g_{*}}{100}\right)^{1/6}.
\end{equation}
This simple picture has been recently challenged in~\cite{Bodeker:2017cim}, where it is shown that in a first-order EWPT bubbles surrounded by a radiation bath tend instead to reach a limit $\gamma$-factor, even when the plasma is significantly diluted. This is due to \emph{transition radiation} of particles in the bath: as they cross the wall, they can emit vector bosons whose mass changes across the wall. This phenomenon translates into the following friction pressure
\begin{equation}
\label{eq:friction}
\Delta P_{\text{NLO}}\sim \gamma g_{\text{EW}}^{2}\Delta m T^{3}_e,
\end{equation}
where $g_{\text{EW}}$ is the electroweak coupling and $\Delta m$ is the typical difference in masses across the wall.
In  a  first-order EWPT this is $\Delta m\sim m_W$. 
As the bubble expand, $\gamma$ increases till it reaches a critical $\gamma$ value, $\gamma_c$,
at which  $\Delta P_{\text{NLO}}$ balances the  energy difference between the true and false vacua,
and the bubble velocity becomes constant. 
For  a  first-order EWPT  with  $T_e\sim m_W$, one finds $\gamma_c\gtrsim 1/(g^2_{\rm EW})$. 

Now, whenever  $\gamma_c<\gamma_{\rm max}$, 
the bubble  can reach $\gamma$-factors of order $\gamma_c$ before they collide.
If this is the case (as for example for the first-order EWPT), 
most of the energy available in the phase transition  will be    released to the surrounding plasma.
Since this latter also sources gravitational radiation, either from  sound waves or turbulence effects,
the gravitational wave signal will be significantly changed~(see e.g.~\cite{Caprini:2015zlo}). 

\begin{figure}[t]
    \centering
    \includegraphics[width=.8\textwidth]{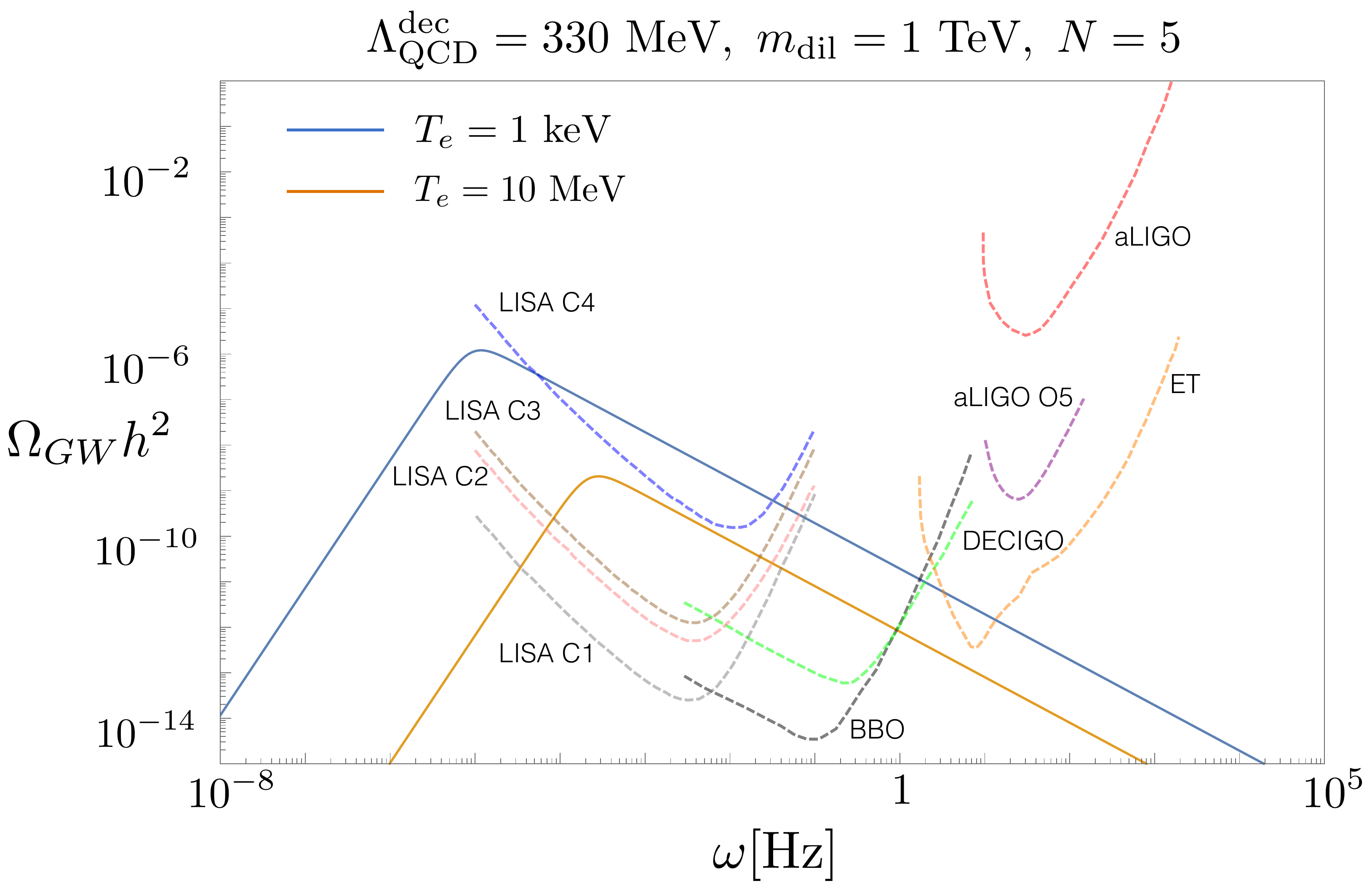}
	\caption{\it Relic abundance of gravitational waves from the collision of cosmic bubbles in the supercooling scenario (thick lines, as in the legend), according to \eq{eq:gwbubbles}. The sensitivity curves of several future ground- and space-based interferometers are also shown (see~\cite{Caprini:2015zlo} for the sensitivity of different configurations of LISA).}
    \label{fig:gwbubbles}
\end{figure}

The confinement phase transition studied in this paper is different from a first-order EWPT studied in~\cite{Bodeker:2017cim}. In particular, the plasma surrounding the bubbles is strongly interacting in our case. Nevertheless, we can expect 
that transition radiation could also  occur  here.
If we assume that \eq{eq:friction} approximately holds in our case, we can estimate
\be
\gamma_c\sim \left(\frac{{\rm TeV}}{T_e}\right)^3\,,
\ee
which is larger than $\gamma_{\rm max}$ for $T_e\lesssim 100$ MeV, indicating that $\gamma_c$ cannot be achieved for 
these small values of $T_e$.
As we have seen in the previous section, in the  long supercooling scenarios  considered here, we can easily achieve  
$T_e\lesssim 100$ MeV (see for example  Fig.~\ref{efolds}).
  If this is the case,
  bubbles would be effectively in the runaway regime until collision. 
We can then envision the following options for the gravitational wave signal in our scenarios:
\begin{itemize}
\item[$\mathbf{1}$]{$T_{e}\gg 100~\text{MeV}$: bubbles expand and reach  $\gamma_c$, giving a 
 gravitational wave signal  dominantly sourced by the plasma which surrounds the bubbles. The
    amplitude and spectrum of this type of signals were analyzed  in~\cite{Caprini:2015zlo},
      but this may not straightforwardly extend to our case where  $\alpha$ is very large. 
      Nevertheless, in the cases which are currently understood, the slope of the spectrum is larger than in \eq{eq:gwbubbles}, the amplitude is linear in $H(T_{\text{RH}})/\beta$ and for $H(T_{\text{RH}})/\beta \sim 1$ it is of the same order of magnitude as \eq{eq:gwbubbles}.}
\item[$\mathbf{2}$]{$T_{e}\lesssim 100~\text{MeV}$: in this case it is reasonable to assume that bubbles effectively run away and the effects of the plasma can be neglected. The gravitational wave signal is given by \eq{eq:gwbubbles} and depends only on $H(T_{\text{RH}})/\beta$ for large $\alpha$.
As a concrete example, let us  consider the exit induced by QCD with $S_B$ given by \eq{eq:actionqcd}. In this case, we have $\beta/H(T_{\text{RH}})\simeq 100/\ln(\Lambda_{\text{QCD}}^{\text{dec}}/T_{e}) - 4$, according to \eq{eq:beta}. Thus, having $T_{e}\ll \Lambda_{\text{QCD}}^{\text{dec}}$ leads to $\beta/H(T_{\text{RH}})\sim 1-10$. Very interestingly, we therefore find that whenever the friction of the plasma can be neglected by exiting from supercooling at sufficiently low temperatures, the duration of the phase transition is very long and the GW signal is maximized. This conclusion is not significantly affected by the exact value of $m_{\text{dil}}$ and $N$. This is true also for $\Lambda_{\text{QCD}}^{\text{dec}}$, as long as $\Lambda_{\text{QCD}}^{\text{dec}}\gtrsim 100~\text{MeV}$.\footnote{If $\Lambda_{\text{QCD}}^{\text{dec}}\ll100~\text{MeV}$ and exit occurs due to QCD, then friction can always be neglected, independently from the value of $\beta/H(T_{\text{RH}})$.} 

In Fig.~\ref{fig:gwbubbles} we show two representative gravitational wave spectra, obtained by fixing $m_{\text{dil}}=1~\text{TeV}, \Lambda_{	\text{QCD}}^{\text{dec}}=330~\text{MeV}, N=5$ and $|\tilde{\lambda}|=0$ for simplicity. We vary instead the exit temperature $T_{e}$. We see that a very strong signal can indeed be obtained if exit occurs much below $\Lambda_{\text{QCD}}^{\text{dec}}$ (blue curve): the peak amplitude is close to the upper bound from dark radiation at BBN~\cite{Caprini:2018mtu} and the signal is detectable at the least sensitive configuration of the LISA interferometer, possibly even at the ET. A weaker signal (orange line), which is nevertheless detectable at more sensitive configurations of LISA, is instead obtained if the exit occurs closer to $\Lambda_{\text{QCD}}^{\text{dec}}$. Both signals are significantly larger than the more commonly considered case $\beta/H(T_{\text{RH}})\gtrsim 100$. However, we stress that a detailed analysis of the thermal friction in the confinement phase transition is required to more confidently assess the shape and size of the GW spectrum in our case.}

\end{itemize}

\section{Conclusions}
\label{conclusions}

We have studied  the conditions  for having a long period of supercooling driven by strongly-coupled theories at the TeV,
which arises due to the small tunneling rate of the confinement phase transition.
We have calculated these rates for several examples, and showed that quite generically (for reasonable values of $N$ and the dilaton mass) the universe gets trapped in the deconfined phase, starting a new period of inflation.  
As the supercooled universe expands and cools down, the temperature reaches the QCD strong scale, $\lqcddec$,
where  QCD effects become relevant.
We have shown that these effects can be calculated for $T\gtrsim \lqcddec$ but  are suppressed 
by $N_c^2/N^2$.  For  $T< \lqcddec$ the CFT drastically changes as 
 QCD-color states get massive and decouple from the rest of the CFT,
affecting the   tunneling rate  by  $O(1)$ effects.
Under  some reasonable assumptions we have shown that these can lead to an end of  supercooling   
after $N_e\sim 5-10$ efoldings.
 
 We have also studied the cosmological consequences of a long period of supercooling. 
Very generically, at the end of supercooling  any DM candidate is diluted by \eq{eq:entropy}.
Under the assumption  that DM is not reheated again after the phase transition,
we have seen that  WIMP-like DM candidates need to have  very different properties.
In particular, see \ref{eq:relicdm}, their masses can be  as large as $10^9-10^3$ GeV for generic DM couplings
of order $1-10^{-3}$ for $T_e\sim$ 100 MeV.
For DM candidates that get masses from the TeV strong scale   $m_\chi\sim \lc$, and are 
massless in the deconfined phase, their relic abundances are mainly   determined by 
their thermal number density  at $T\sim T_c$ and the dilution factor \eq{eq:entropy}, as shown in \eq{eq:relicmassless}. This can also be the case for baryons, as we have seen in the particular case  in which the baryon asymmetry is generated  from  decaying particles  that were in thermal equilibrium at high temperatures.
This  provides an interesting explanation for having a ratio of baryon to DM energy densities roughly
proportional to the ratio of their masses, even though  the mechanisms that originate them were  very different.

One of the main impact of a long supercooling epoch is for the QCD  axion abundance.
We find that during supercooling we can  have  $H\sim m_a$ (not possible for inflation at higher energies),
implying that the axion can oscillate towards its minimum.
This is possible in models in which the EW symmetry is broken in the deconfined phase.
This leads to  an extra dilution of   axions as compared to the standard case, 
which changes the predictions for being DM.
We find  that larger values of $F_a$ are  required (see Fig.~\ref{fig:nspace}),
weakening the signal in experiments searching for DM axions such as ADMX~\cite{Du:2018uak}.
On the other hand,  the contributions from axionic topological defects can be enhanced in a supercooled universe. 
This is because, due to the inflationary expansion, the axionic string-wall network can collapse later than in the standard cosmology. Therefore, the network can provide a larger contribution to the axion relic abundance, which would limit the allowed values of $F_a$, see Fig.~\ref{fig:walls}.

Finally, we have also explored the GW  signal arising from the end of the supercooling epoch. We identified two peculiarities of our setup, which can simplify the estimates of the GW spectrum, as well as maximize the signal. Firstly, due to the long inflationary expansion, the plasma which surrounds the bubble can be severely diluted: this implies $\alpha\gg 1$ and that the friction on the bubble motion can be likely neglected. Therefore, bubble collisions can be effectively considered as the main source of gravitational radiation. Secondly, due to the very same reason, one can have $\beta/H\sim O(1)-O(10)$, which leads to a stronger GW signal than in the usual case, see Fig.~\ref{fig:gwbubbles}.

There are many things left for future work.
It would be very useful   to better understand  how the CFT behaves at $T<\lqcddec$ in order
to get a better prediction for $T_e$.
For DM candidates we have seen that the  reheating process at the end of the phase transition
deserves more attention as it is crucial to  determine their abundances.
Also it is important to understand  transition radiation effects in the confinement phase transition
as this plays a crucial role in the GW signal.

\medskip 
\section*{Acknowledgments}
We thank J. Garriga., B. von Harling, A. Katz, O. Pujol\`as and G. Servant for discussions.
A.P. has been  supported by the Catalan ICREA Academia Program.
This work has  also been partly supported by the grants FPA2014-55613-P, FPA2017-88915-P, 2017-SGR-1069 and SEV-2016-0588.

\appendix

\section{Holography and Supercooling}
\label{appendixA}

Using the AdS/CFT dictionary \cite{Maldacena:1997re,Witten:1998qj}, we can identify weakly-coupled five-dimensional models with the same properties of the strongly-coupled theories described in Section \ref{SIS}. 
This provides us with simple calculable models to address the supercooling phenomena in a more quantitative way.
Many examples were considered before in the literature
\cite{Creminelli:2001th,Randall:2006py, Kaplan:2006yi, Nardini:2007me, Hassanain:2007js, Konstandin:2010cd, Konstandin:2011ds, Konstandin:2011dr, Bunk:2017fic, Dillon:2017ctw, Megias:2018sxv}. 
Here we briefly recall the entries of the dictionary that are useful for studying our problem.

The ``dual'' five-dimensional models  are characterized  by an AdS$_5$ geometry: 
\begin{equation}
ds^2=\frac{L^2}{z^2}\big(\eta_{\mu\nu}dx^\mu dx^\nu+dz^2\big)\, ,
\label{ads5}
\end{equation}
solution to the 5D Einstein equations for metric $G_{MN}$ with negative cosmological constant
\be
S_5=2M_5^3\int d^5 X \sqrt{-G}(\mathcal{R}_5(G)+12L^{-2})\,,
\label{einstein}
\ee
where $L$ is the AdS curvature radius, $M_5$ the 5D Planck mass and $\mathcal{R}_5$ the 5D Ricci scalar. 
This five-dimensional space is  assumed to be ending  by a hard-wall, often called IR-brane, at some position in the extra dimension, $z=\zir$. For \eq{ads5} to actually be a solution ending with a hard-wall at $z_{\rm IR}$, the hard-wall must have a negative tension $\tau_{\rm IR}=-24M_5^3 L^{-1}$. When this condition is met the value of $z_{\rm IR}$ is not dynamically fixed
but  free to take any value. In fact $z_{\rm IR}$  spans a moduli space of equivalently solutions to \eq{einstein}. 
The AdS/CFT dictionary tells that the radion  corresponds to the  dilaton of the CFT  model via  the identification $\mu=1/\zir$.
The tuning of the IR-brane tension  $\tau_{\rm IR}$ to the value above
 corresponds to tune  the dilaton potential to zero.  
  

The moduli space parametrized by $z_{\rm IR}$ is actually part of a larger manifold of solutions, parametrized by $z_{\rm IR}$ itself and a four dimensional constant matrix ${g}_{\mu\nu}$, generalizing the matrix $\eta_{\mu\nu}$ in \eq{ads5}. This moduli space is associated to massless fields in 4D, corresponding to the spacetime dependent fluctuations of the moduli ``coordinates'' $z_{\rm IR}$ (or $\mu$) and ${g}_{\mu\nu}$. When rewriting the Einstein 5D action \eq{einstein} in terms of the moduli fields, we get the 4D action
\be
S= (M_5 L)^3 \int d^4 x \sqrt{-g} \left(2(L^{-2}-\mu^2)\mathcal{R}(g)-12(\partial_\mu\, \mu(x))^2\right)\,,
\label{holokin}
\ee
where $\mathcal{R}$ is the 4D Ricci scalar constructed from $g_{\mu\nu}(x)$. The AdS/CFT correspondence relates 
  the 5D Planck mass in units of  $1/L$ to the rank of the  4D gauge  group $N$, according to  $(M_5 L)^3\approx N^2/16\pi^2$. The dilaton $\mu$ is not canonically normalized, and \eq{holokin}
   fixes the coefficient of \eq{dilkin} to $c_1=12$.
The massive  Kaluza-Klein modes are interpreted in the 4D as resonances of the strongly coupled CFT, and their masses are given by
\be
m_n\approx (n+\frac{1}{4})\pi \mu\ ,\ \ \ n=1,2,...\,,
\ee
corresponding to fix $r_i\sim \pi$ in \eq{massr}.

In the 5D models discussed here we also  need the equivalent of the  4D marginally relevant  operator \eq{deformation} with  ${\rm Dim}[{\cal O}_g]=4-\epsilon$.
By use of  the  AdS/CFT dictionary, we know that  operators ${\cal O}_\Phi$ in the CFT 
are mapped to 5D fields in AdS with the same quantum numbers. Also the
 dimensions of the 4D scalar operators  are related to the  masses of the 5D fields according to
\be
{\rm Dim}[{\cal O}_\Phi]=2+\sqrt{4+M^2_{\Phi} L^2}\, ,
\label{dictionary}
\ee
where $M_\Phi$ is the mass of $\Phi$. 
Therefore, 
the  marginally relevant  operator \eq{deformation}
 is associated to an almost massless scalar in the five-dimensional theory, $M^2_\Phi\propto \epsilon$. 
Furthermore,  we have the mapping  $g=\Phi|_{z_{\rm UV}}\not=0$, that triggers a nonzero profile for $\Phi$ in the bulk.

Let us now move to the symmetry breaking pattern.
The AdS/CFT dictionary tells that 
having the global symmetries  $\cal G$ in the CFT corresponds   to   gauging a group $\cal G$ 
in the 5D bulk, and the SSB $\cal G\to H$ by a scalar condensate translates to
having an extra 5D scalar in the bulk, $H_5$, transforming under $\cal G$, that gets an  nonzero VEV on the IR-brane.
The SM Higgs $H$ is associated to be one of the zero-modes of $H_5$ after KK-reduction.

\begin{figure}[t]
\centering
\includegraphics[width=0.43\textwidth]{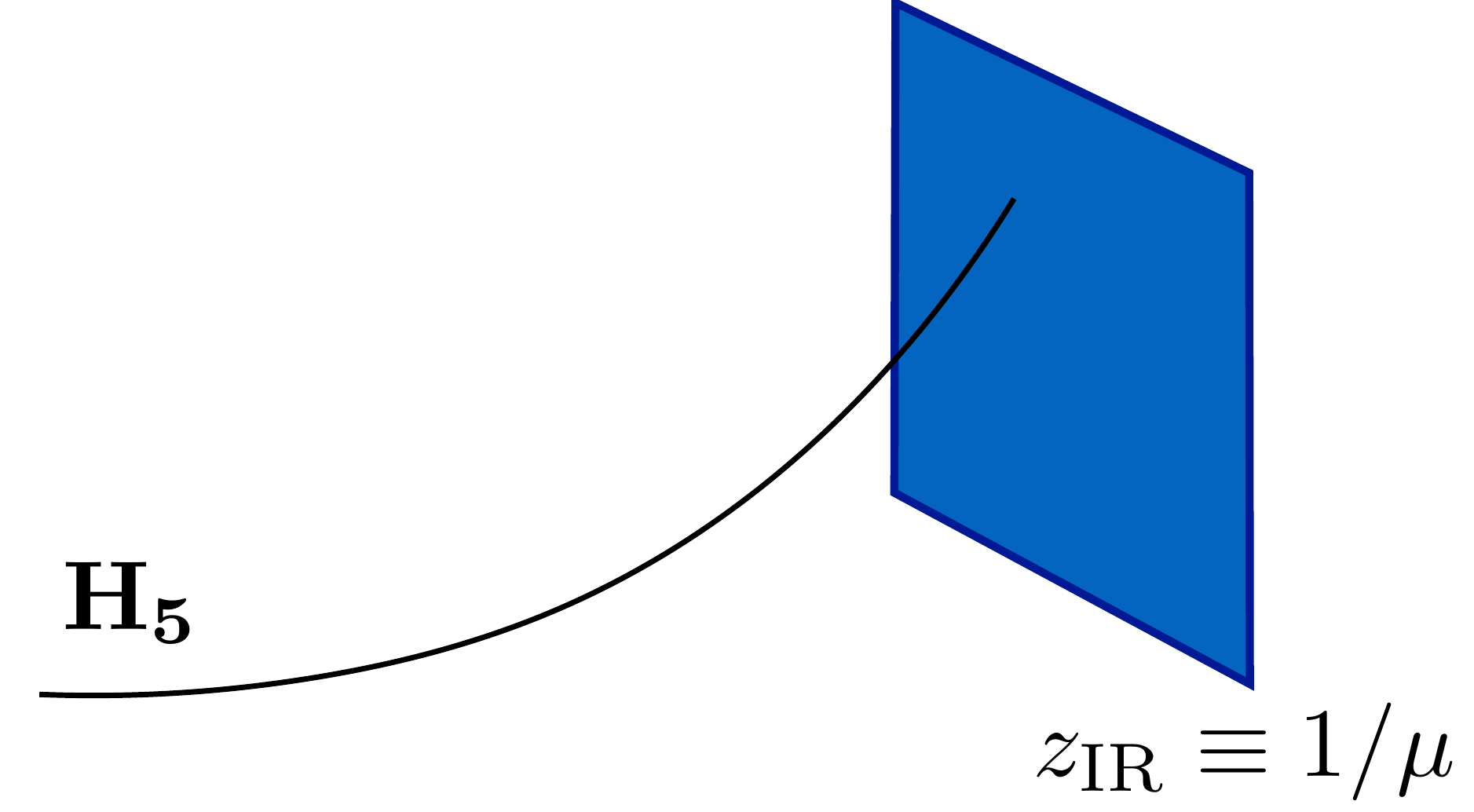}\ \ \ \ \ \ \ \ \ \ 
\includegraphics[width=0.48\textwidth]{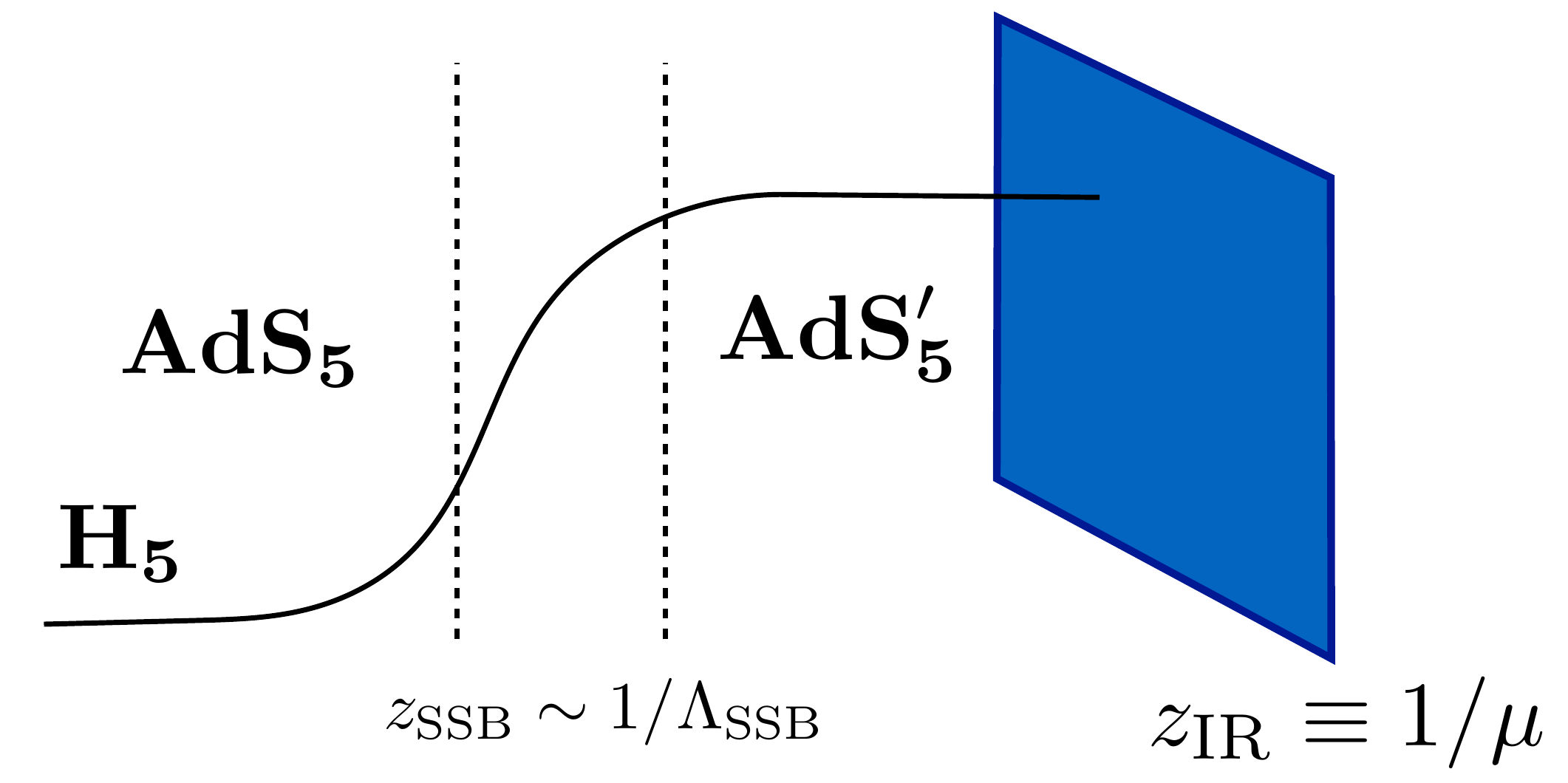}
\caption{\it  Models of type \I (left) and type \II (right).}
\label{modelplots}
\end{figure}

In  realistic models the potential of the radion/dilaton should not be fixed to zero 
but be generated from the 5D model in order  to naturally make $\langle\mu\rangle\sim {\rm TeV}$.
Let us  qualitatively discuss how this can arise  in the 5D versions of models of type \I and \II studied in the paper:

\begin{enumerate}[I.]
\item 
In this case the radion/dilaton  potential arises from the almost massless field $\Phi$,
as in  Goldberger-Wise models   \cite{Goldberger:1999uk}.
$H_5$ has a potential on the IR-brane that gives a non-zero VEV for $H_5$ at the IR-brane.
We have in this case that the VEV profile of $H_5$ is peaked towards the IR-brane and therefore 
$\langle H\rangle\propto\mu$.
This  possibility is illustrated by the left figure of Fig.~\ref{modelplots}.
\item 
$H_5$ is coupled to $\Phi$ that gives him  a $z$-dependent mass, $M^2_{H_5}\sim \Phi(z)$.
At some  $z$, we have that   $M^2_{H_5}$ becomes smaller than 
the BF-bound ($M^2_{H_5}<-4/L^2$), and 
$H_5$ becomes  an  AdS$_5$ tachyon. 
In this case  $H_5$ gets  a nonzero profile in the 5D bulk that  grows as $z^2$, till
 $H_5$ reaches  the  minimum of  its 5D potential $V(H_5)$ and  becomes constant in $z$. 
 The value of $z$ where the maximum is achieved, $z_{\rm SSB}$,  is associated with 
 the SSB scale $\lssb\sim 1/z_{\rm SSB}$.
  When including the metric back-reaction,  one gets that the AdS$_5$ geometry evolves at around  $z_{\rm SSB}$ into another 
AdS$_5$ with different curvature. The dual 4D interpretation of this 
 is that a CFT  RG-evolves  into another CFT at $\lssb$.
The tachyon profile  gives also a contribution to the  radion potential and a minimum for $\mu$ can be achieved, as studied in Ref.~\cite{inpreparation}. One usually finds $\langle\mu\rangle\sim \lssb$.
When moving the IR-brane off-shell towards the horizon $z\to \infty$,
the VEV of $H_5$ is not proportional to $\mu$ and then does not go to zero as in  type \I models.
This is illustrated in the right figure  of Fig.~\ref{modelplots}.
\end{enumerate}

\noindent Let us  finally provide  the 5D  implementation  of \eq{portal}. The fermionic operators ${\cal O}_{L,R}$
corresponds to  5D fermions with  the same quantum numbers as the SM quarks. 
This implies that the QCD group must be  a gauged symmetry in the 5D bulk.
At low energies, the  zero-modes of these 5D fermions corresponds to the SM quarks.

\subsection{The deconfined phase}\label{appdec}
In 5D holographic versions  the deconfined phase is characterized by a  different solution to the 5D Einstein equations (arising  from \eq{einstein}) at \textit{finite temperature} $T$  where we must take the  Euclidian version of the theory, with compact time $it\in [0,1/T]$.
This solution is called AdS$_5$-Schwarzschild Black Hole (AdS-BH) geometry, and the metric can be expressed as
\begin{equation}
ds^2=-\frac{L^2}{z^2}\left(-f(z)dt^2+dx_i^2+\frac{1}{f(z)}dz^2\right)\ ,\ \ \  \
f(z)=1-\left(\frac{z}{z_h}\right)^4\, .
\label{bh}
\end{equation}
This metric has a BH horizon at $z_h$ replacing the infrared brane  ($z$ runs from $z=0$ to  $z=z_h$). The  BH horizon is associated to a Hawking temperature, given by $T_h=1/(\pi z_h)$.
To avoid singularities, we must match the BH Hawking temperature with the  temperature of the system: $T=1/(\pi z_h)$ (see \ref{appfree} for more details).

On the AdS-BH background \eq{bh} the 5D fields which were discussed above, and in particular $H_5$, take in general a different profile. We are  especially interested  in  the value of  $\lssb$  in models \I and  \II.
In type  \I  models  $H_5$ 
 cannot get a nonzero profile in  the AdS-BH solution 
 since, contrary to the IR-brane, there is no potential for $H_5$ in the BH solution.
On the other hand, in  models of type \II, $H_5$ 
 is an AdS$_5$ tachyon and then  also turns  on 
  even in the absence of an  IR-brane potential. Indeed, we can find a nonzero 
profile for  $H_5$ in the AdS-BH where at $z=z_h$ the regularity of the solution implies
\be
\left.\frac{4}{z}\partial_z H_{5}+V(H_{5})\right|_{z=z_h}=0\,.
\ee
Therefore,  the EW symmetry is broken  in type \II models    also in the deconfined phase.

\subsection{Free energy of the deconfined phase}\label{appfree}

The energy of solution \eq{bh} is interpreted in the 4D as the free energy of the plasma phase, with identification $z_h=1/(\pi T_{loc})$. In fact, when $z_h\neq 1/(\pi T)$ the metric \eq{bh} has a conical singularity at  $z=z_h$ that  gives a finite contribution to the free energy density  equal to \cite{Creminelli:2001th}
\be
\mathcal{F}_{\rm cone}=8\pi^4(M_5 L)^3 T_{loc}^4\left(1-\frac{T}{T_{loc}}\right)\,.
\label{cone}
\ee
We see that it correctly vanishes when $T_{loc}=T$. A remaining piece comes from the regular region $z\neq z_h$. This is UV divergent, but what matters for tunneling is its finite difference with the energy of full AdS$_5$.  
This gives
\be
\mathcal{F}_{\rm BH}(z\neq z_h)-\mathcal{F}_{\rm AdS}=-2\pi^4 (M_5 L)^3 T_{loc}^4.
\label{regular}
\ee
When added together, the two contributions \eq{cone} and \eq{regular} give
\be
\mathcal{F}_{\rm BH}=2\pi^4(M_5 L)^3(3T_{loc}^4-4T T_{loc}^3)=\frac{\pi^2}{8}N^2(3T_{loc}^4-4T T_{loc}^3)\,,
\label{fbh}
\ee
where the last equality comes from the identification $(M_5 L)^3=N^2/16\pi^2$. With this identification 
we get  $c_2=\pi^2/8$  for  \eq{fdec}. The minimum of \eq{fbh} is attained when $T=T_{loc}$, as it must be since Einstein equations forbid conical singularities.

This result can also be determined, modulo an overall constant, from elementary thermodynamics and scale invariance  as follows. If we put a small region of volume $V$ to temperature $T_{loc}$, its free energy density (with respect to a thermal bath at temperature $T$) will be $\mathcal{F}=(E(T_{loc},V)-T S(T_{loc},V))/V$. The total energy and entropy $E$ and $S$ are extensive variables, and scale invariance tells that the densities $E/V$ and $S/V$ are appropriate powers of the only available scale $T_{loc}$ (that is respectively $T_{loc}^4$ and $T_{loc}^3$)
\be
\mathcal{F}(T_{loc};T)=a T_{loc}^4-b T T_{loc}^3\,.
\ee
We know that $\mathcal{F}$ is a \textit{minimum} at thermal equilibrium, that is when $T_{loc}=T$. This fixes the ratio $a/b=3/4$ universally for \textit{any} CFT. 

\section{Axion dark matter from string-wall network}
\label{app:topological}

The aim of this appendix is to review the estimate of the contribution to the total axion dark matter abundance from the annihilation of the string-wall network (see e.g.~\cite{Kawasaki:2014sqa}). The energy of the network is dominated by the walls, thus we have 
\begin{equation}
\rho_{\text{network}}\simeq \sigma H,
\end{equation}
where $\sigma\simeq 8 m_{a}(0) F_{a}^{2}$ is the wall tension. 
This energy is eventually entirely released in axion quanta which behave like dark matter. The process is expected to occur very quickly after $T_{\star}$. The relic number density of the radiated particles is thus given by
\begin{equation}
n_{\text{network}}(t_{0})=\frac{\rho_{\text{network}}}{\bar{E}_{a}}\left(\frac{a(t_{\star})}{a(t_{0})}\right)^{3},
\end{equation}
where $\bar{E}_{a}$ is the mean energy of the radiated axions at $T_{\star}$. This has to be extracted from numerical simulations: here we follow the results of~\cite{Kawasaki:2014sqa}, according to which $\bar{E}_{a}\simeq 3.23~m_{a}(T_{\star})$. The relic energy density of such axions is given by
\begin{equation}
\rho_{\text{network}}(t_{0})=\frac{m_{a}(o)}{3.23 m_{a}(T_{\star})}\rho_{\text{network}}(T_{\star})\left(\frac{a(t_{\star})}{a(t_{0})}\right)^{3}.
\end{equation}
Finally the relic abundance is
\begin{equation}
\Omega_{\text{network}}h^{2}\simeq 0.03 \left(\frac{61.75}{g_{*}(T_{\star})}\right)^{1/4}\left(\frac{F_{a}}{10^{10}~\text{GeV}}\right)\left(\frac{100~\text{MeV}}{T_{\star}}\right).
\end{equation}

\providecommand{\href}[2]{#2}\begingroup\raggedright\endgroup

\end{document}